\documentstyle[12pt]{article}
\title{Toward verification of the Riemann hypothesis:  Application of
the Li criterion}
\author{Mark W. Coffey\\
Department of Physics\\
Colorado School of Mines\\
Golden, CO  80401\\
(Received $\mbox{~~~~~~~~~~~~~~~~~~~~~~~~~~~~~~~2004}$)}
\date{August 10, 2004}   
\begin{document}
\maketitle
\baselineskip=25 pt
\begin{abstract}

We substantially apply the Li criterion for the Riemann hypothesis to
hold.  Based upon a series representation for the sequence $\{\lambda_k\}$,
which are certain logarithmic derivatives of the Riemann xi function 
evaluated at unity, we determine new bounds for relevant Riemann zeta 
function sums and the sequence itself.  We find that the Riemann hypothesis 
holds if certain conjectured properties of a sequence $\eta_j$ are valid.  
The constants $\eta_j$ enter the Laurent expansion of the logarithmic 
derivative of the zeta function about $s=1$ and appear to have remarkable 
characteristics.  {\em On our conjecture}, not only does the Riemann 
hypothesis follow, but an inequality governing the values $\lambda_n$ and
inequalities for the sums of reciprocal powers of the nontrivial zeros of
the zeta function.

\end{abstract}
 
\vspace{.25cm}
\baselineskip=15pt
\centerline{\bf Key words and phrases}
\medskip 

\noindent
Riemann zeta function, Riemann xi function, logarithmic derivatives, 
Riemann hypothesis, Li criterion, Stieltjes constants, Laurent expansion,
L function, Dirichlet series, Hecke L function, Dedekind zeta function,
extended Riemann hypothesis

 
\baselineskip=25pt
\pagebreak
\medskip
\centerline{\bf Introduction}
\medskip


In this paper, we reduce the verification of the Riemann hypothesis to a
conjecture concerning the behavior of certain coefficients $\eta_k$ which
appear in the Laurent expansion of the logarithmic derivative of the zeta
function about $s=1$.  Moreover, should the conjectured property hold, we 
would then derive a result stronger than the Riemann hypothesis itself.  Namely, 
we would have obtained an explicit lower bound for a sequence $\{\lambda_j\}$ of 
certain logarithmic derivatives of the xi function evaluated at unit argument.  
Since the coefficients $\eta_j$ can be written as particular limits involving 
the von Mangoldt function $\Lambda$, these quantities seem to encapsulate both
number theoretic and analytic information.  Such information is expected to
be encountered in any rigorous denial or verification of the Riemann
hypothesis.

We first present an overview of our approach.  We shall use the Li equivalence
for the Riemann hypothesis to hold \cite{li}.  This equivalence results from 
a necessary and sufficient condition that the logarithmic derivative of the
function $\xi[1/(1-z)]$ be analytic in the unit disk.  
The function $\xi$ is determined from the Riemann zeta function $\zeta$
by way of the relation $\xi(s)=(s/2)(s-1)\pi^{-s/2}\Gamma(s/2)\zeta(s)$,
where $\Gamma$ is the Gamma function 
\cite{edwards,ivic,karatsuba,titch,riemann}.  
Then the xi function satisfies the functional equation $\xi(s)=\xi(1-s)$.  
The Li equivalence
states that a necessary and sufficient condition for the nontrivial zeros of
the Riemann zeta function to lie on the critical line Re $s=1/2$ is that
a sequence of real numbers $\lambda_k$ is nonnegative for every integer $k$.
In this paper, we investigate the behavior of this sequence, based upon a
series representation previously derived \cite{coffey}.  Of this representation,
we are able to bound one finite sum and make progress in bounding the other.
We find that indeed the sequence $\{\lambda_k\}$ has only nonnegative numbers, 
subject to the conjectured properties of the sequence $\eta_j$.  Among other 
connections, the $\eta_j$ can be readily related to the Stieltjes constants 
$\gamma_k$.  This gives one of several avenues for further investigation of the 
$\eta_j$ expansion coefficients.  This paper includes a series of appendices
A--L which contain various extensions of our summation estimations, occasional
alternative proofs, series representations analogous to that for $\lambda_n$
relevant to other L-functions, tabulated numerical values, derivative relations 
of the Riemann zeta function, alternative expansions of the logarithmic derivative
of the Riemann zeta and xi functions, remarks on the integer-order derivatives of
the Dedekind zeta and xi functions, and other reference material.  

We stress that we do not just conjecture as to the nature of the $\eta_j$'s, 
but provide a perhaps strong plausibility argument in favor.  In 
addition, current numerical evidence \cite{krzysztof} seems to fully support 
our conjecture.

The sequence $\{\lambda_n\}_{n=1}^\infty$ is defined by
$$\lambda_n = {1 \over {(n-1)!}}{d^n \over {ds^n}}[s^{n-1}\ln \xi(s)]_{s=1}.
\eqno(1)$$
Then Li's criterion for the Riemann hypothesis to hold is that all 
$\{\lambda_n\}_{n=1}^\infty$ are nonnegative \cite{li}.  We note that Li's
convention for the xi function has a factor of two difference:  $\xi_{Li}(s)
=2\xi(s)$, although this is immaterial in logarithmic derivatives such as
$\lambda_j$.  We also have 
$$\lambda_n = {{(-1)^n} \over {(n-1)!}}{d^n \over {ds^n}}[(1-s)^{n-1}\ln 
\xi(s)]_{s=0}, \eqno(2)$$
and $\xi(0)=-\zeta(0)=1/2$.  (Hence $\xi_{Li}(0)=1$.)  The approximate 
numerical values for the first few $\lambda_j$'s are:  $\lambda_1 \simeq 
0.0230957$,  $\lambda_2 \simeq 0.0923457$, and $\lambda_3 \simeq 0.207639$
(see Appendix D).
In fact, we have $\lambda_1=-B$, where $B \equiv \ln 2+(1/2)\ln \pi-1-\gamma/2
\simeq -0.0230957$, and $\gamma \simeq 0.5772156649$ is the Euler constant. 
This follows from the logarithmic derivative \cite{ivic}
$${{\xi'(s)} \over {\xi(s)}}=B+\sum_\rho\left ({1 \over {s-\rho}} + {1 \over
\rho}\right ), \eqno(3)$$
where $\rho$ runs over all the nontrivial zeros of the zeta function.  Thus
$\xi'(s)/\xi(s)=\sum_\rho (s-\rho)^{-1}$, which is consistent with
$\xi_{Li}(s)=\prod_\rho(1-s/\rho)$.  In general, the $\lambda_j$'s are 
connected to sums over the nontrivial zeros of $\zeta(s)$ by way of \cite{li}
$$\lambda_n=\sum_\rho\left [1-\left(1-{1 \over \rho}\right)^n \right ].  
\eqno(4)$$          

By using the Laurent expansion of the zeta function about $s=1$,
$$\zeta(s)={1 \over {s-1}}+ \sum_{n=0}^\infty {{(-1)^n} \over {n!}}\gamma_n
(s-1)^n, \eqno(5)$$
where the $\gamma_k$ are the Stieltjes constants \cite{ivic,scref}, it is 
possible to write a closed form for the $\lambda_j$'s. The Stieltjes 
constants can be evaluated from the expression
$$\gamma_k=\lim_{N \to \infty}\left (\sum_{m=1}^N {1 \over m}\ln^k m-
{{\ln^{k+1} N} \over {k+1}}\right ), \eqno(6)$$
and several other forms have been given \cite{scref}.  For instance, we have
$$\lambda_2=1+\gamma-\gamma^2+\pi^2/8-2\ln 2-\ln \pi-2\gamma_1, \eqno(7)$$
and
$$\lambda_3={1 \over 2}\left [2+{3 \over 4}\pi^2-6\ln 2-3\ln \pi-12\gamma_1
+\gamma[3+2(\gamma-3)\gamma+6\gamma_1]+3\gamma_2-{7 \over 4}\zeta(3)
\right ].  \eqno(8)$$
It is not difficult to determine that $\lambda_j$ contains the term
$-(-1)^j [j/(j-1)!]\gamma_{j-1}$.     
The successive $\lambda_j$'s can be related in several different ways,
including simply
$$\lambda_{n+1}=\lambda_n+{1 \over {n!}}\left[{d^n \over {ds^n}}s^n {{\xi'(s)}
\over {\xi(s)}}\right ]_{s=1}.  \eqno(9)$$
Alternatively, in the particular case of $\lambda_2$, one can write
$\lambda_2=2\lambda_1-\lambda_1^2+\lim_{s\to 1}s\xi''(s)/\xi(s)$.
Elsewhere, we have very recently obtained a general explicit relation for
$\lambda_k$ in terms of the Stieltjes constants \cite{coffeygamma}.

\medskip
\centerline{\bf Alternative representation of Li's $\lambda_j$'s}

Of particular importance for the purposes of this paper is an alternative 
formula for the particular sequence of logarithmic derivatives of the 
Riemann xi function given in Eq. (1).  Due to the centrality of this result, 
we briefly review the derivation \cite{coffey} of the following representation,
\newline{\bf Theorem 1}
$$\lambda_n=-\sum_{m=1}^n {n \choose m} \eta_{m-1}+\sum_{m=2}^n
(-1)^m {n \choose m} (1-2^{-m})\zeta(m)+1-{n \over 2}(\gamma
+\ln \pi + 2\ln 2), \eqno(10)$$
where the constants $\eta_j$ can be written as
$$\eta_k={(-1)^k \over {k!}} \lim_{N \to \infty}\left (\sum_{m=1}^N {1 \over 
m} \Lambda(m)\ln^k m - {{\ln^{k+1} N} \over {k+1}}\right ), \eqno(11)$$
and $\Lambda$ is the von Mangoldt function 
\cite{edwards,ivic,karatsuba,titch,riemann,ivic93}.
From the expansion around $s=1$ of the logarithmic derivative of the
zeta function, 
$${{\zeta'(s)} \over {\zeta(s)}}=-(s-1)^{-1}-\sum_{p=0}^\infty \eta_p (s-1)^p, 
\eqno(12)$$ 
we have
$$\ln \zeta(s)=-\ln(s-1)-\sum_{p=1}^\infty {\eta_{p-1} \over p}(s-1)^p,
\eqno(13)$$
giving
$$\ln \xi(s)=-\ln 2 +\ln s-{s \over 2}\ln \pi+\ln \Gamma(s/2)
-\sum_{p=1}^\infty {\eta_{p-1} \over p}(s-1)^p.  \eqno(14)$$
The radius of convergence of the expansion (12) is $3$, as the first
singularity encountered is the trivial zero of $\zeta(s)$ at $s=-2$.
We then evaluate
$${d^n \over {ds^n}}[s^{n-1}\ln \xi(s)]_{s=1}=(n-1)!\sum_{j=0}^{n-1}
{n \choose j}{1 \over {(n-j-1)!}}\left [{d^{n-j} \over {ds^{n-j}}}\ln \xi(s) 
\right]_{s=1},  \eqno(15)$$
using in particular the special values
$\psi(1/2)=-\gamma-2 \ln 2$ and
$\psi^{(n)}(1/2)=(-1)^{n+1}n!(2^{n+1}-1)\zeta(n+1)$ for $n \geq 1$, where
$\psi=\Gamma'/\Gamma$ is the digamma function and $\psi^{(j)}$ is the
polygamma function.  Finally, the sum in Eq.\ (15) over $j$ can be
converted to a sum over $m=n-j$ and the simple result
$-\sum_{m=1}^n(-1)^m {n \choose m}=1$ used, yielding
Eq.\ (10).  

The Laurent expansion of $\zeta'/\zeta$ with the form of the constants
$\eta_j$ in Eq. (11) can be developed by applying Theorem 1 of Ref. 
\cite{ivic93}.  In this case, the counting function $A(x)=-\sum_{n \leq x}
\Lambda(n) = -\psi(x)$, where $\psi$ is the Chebyshev function, and the
error term is given by $u(x)=x-\psi(x)$.
Equation (10) has been derived alternatively in Ref. \cite{bombieri} by
a method connected with A. Weil's explicit formula.  Our approach is
independent, and we believe, more direct.  In Appendix G we present an
extension of Theorem 1 which accounts explicitly for the presence of the
first six trivial zeros of the zeta function.

\medskip
\centerline{\bf Estimation of Sums}
\medskip

We characterize each of the two summation terms on the right side of Eq. (10)
in turn.  Before this discussion, we emphasize that the first few $\lambda_j$'s
may be explicitly written, as indicated above, and directly verified to be
positive.  In addition, we note that apparently it is already known 
\cite{biane} that $\lambda_n \geq 0$ for all $n \leq 2.975\ldots \times 10^{17}$.

The sum
$$S_1(n) \equiv \sum_{m=2}^n (-1)^m {n \choose m} (1-2^{-m})\zeta(m),
\eqno(16)$$
may be written in several equivalent ways, including
$$S_1(n)=\sum_{k=0}^\infty\sum_{m=2}^n(-1)^m{n \choose m}{1 \over {(2k+1)^m}}
=\sum_{k=0}^\infty \left [{n \over {2k+1}}-1+{{2^nk^n} \over {(2k+1)^n}}
\right ].  \eqno(17)$$
Equation (17) results from the use of (e.g., \cite{nbs})  
$$(1-2^{-n})\zeta(n)=\sum_{k=0}^\infty (2k+1)^{-n}, ~~~~ n \geq 2, \eqno(18)$$
and interchange of the order of the two summations in Eq. (16).
It appears to be equally profitable to write $S_1$ as
$$S_1(n) = \sum_{k=1}^\infty\sum_{m=2}^n (-1)^m {n \choose m} (1-2^{-m}){1 \over 
k^m}=\sum_{k=1}^\infty\left [{n \over {2k}}+\left(1-{ 1\over k}\right)^n-
\left (1-{1 \over {2k}}\right )^n \right ]. \eqno(19)$$
By using integrals estimating these forms, we obtain 
\newline{\bf Theorem 2}
$$S_1(n) \geq {n \over 2}\ln n + (\gamma - 1){n \over 2} +{1 \over 2}. \eqno(20)$$

Remark.  By inserting an integral representation for $\zeta$ into Eq. (16), it
is possible to obtain the sum $S_1$ in the form
$$S_1(n) = {n \over 2}\int_0^\infty [1+~_1F_1(1-n;2;t/2)-2~_1F_1(1-n;2;t)]
{{dt} \over {(e^t - 1)}}, $$   
where $~_1F_1$ is the Kummer confluent hypergeometric function.  In this equation,
$n ~_1F_1(1-n;2;w)=L_{n-1}^1(w)$, where $L_n^\alpha$ is an associated Laguerre
polynomial.

Proof of Theorem 2.  We proceed on the basis of comparison to the form of
Eq.\ (17), leaving the comparison to Eq.\ (19) to Appendix A.  In addition, we
relegate to Appendix L the application of Euler-Maclaurin summation to $S_1(n)$,
which gives improved estimates of the linear term.

By making a
change of variable in the integral
$$I_1(n) = \int_0^\infty\left [{n \over {2k+1}}-1+{{2^nk^n} \over {(2k+1)^n}}
\right ] dk, \eqno(21)$$
we obtain
$$I_1(n)={1 \over 2}\int_1^\infty \left [{n \over x}-1+{{(x-1)^n} \over x^n}
\right ]dx.  \eqno(22)$$
Another change of variable and an integration by parts yields
$$I_1(n)={n \over 2}\int_0^1[1-(1-t)^{n-1}]{{dt} \over t} -{1 \over 2}(n-1).
\eqno(23)$$
Evaluation of the integration \cite{grad} then gives
$$I_1(n)={n \over 2}[\psi(n)+\gamma -1] + {1 \over 2}.      \eqno(24)$$
Since the integrand of Eq. (21) is monotonically decreasing with $k$,
the inequality (20) readily follows.

{\bf Corollary}  With the other sum term in Eq. (10) denoted as $S_2$, we 
have already obtained the inequality $\lambda_n \geq (n/2)\ln n - (n/2)
(1+\ln \pi+2\ln 2) +3/2 - |S_2|$.  Similarly, it is possible to readily
obtain an upper bound for $S_1$ and $\lambda_n$.  We have, for instance,
$S_1(n) \leq (n/2)\ln n+(\gamma+1)n/2-1/2$.  It is possible to develop
improved coefficients for the linear in $n$ term, giving tighter bounds
for $S_1(n)$, as shown in Appendix L.  

Remarks. (1) Since Li has very recently obtained explicit formulas for both
Dirichlet and Hecke L-functions \cite{li2}, which are analogous to Eq. 
(10), we may expect our summation estimation techniques to aid in making 
progress in verifying the extended and generalized Riemann hypotheses. 
In fact, we may observe that the term $\tau_\chi(n)$ defined in Theorem
1 of Ref. \cite{li2} can be rewritten as
$$\tau_\chi(n)=S_1 - n\ln 2 ~~~~~~~~~~~~~~~~~~~\mbox{if} ~~~~\chi(-1)=1,$$
$$~~~~~~~~=S_0-S_1, ~~~~~~~~~~~~~~~~~~~   \mbox{if} ~~~~\chi(-1)=-1,
\eqno(25)$$
where $S_0$ is defined in Eq. (A.15) of Appendix A.  The second line of
Theorem 2 of Ref. \cite{li2} for $\lambda_E(n)$ may be written as
$$2\left ( 1 - {1 \over 3}\ln 2 \right )n + S_3 - S_0 + 2n + (-1)^n - 1,
\eqno(26)$$
where $S_3$ is defined in Eq. (A.18) of Appendix A.
Inequality results based upon the use of Eqs. (25) and (26) and the
estimations of Appendix A are given in Appendix E, while estimations of
generalized Riemann zeta functions sums are presented in Appendix F.

(2)  In Ref. \cite{voros}, the rate of growth of the sum $S_1(n)$ 
is conjectured.  An integral expression for the Li constants is written and 
a saddle point method is applied.  However, the dominant $O(n\ln n)$ behaviour 
that we have demonstrated is not rigorously obtained.

It is not difficult to show that the following recursion relation
exists between the Stieltjes constants of Eq. (5) and the coefficients
$\eta_j$.  We have $\eta_0=-\gamma_0=-\gamma$ and
$$\eta_n=-(-1)^n \left [{{(n+1)} \over {n!}} \gamma_n + \sum_{k=0}^{n-1} 
{{(-1)^{k-1}} \over {(n-k-1)!}} \eta_k \gamma_{n-k-1} \right ].  \eqno(27)$$
This equation is equivalent to the statement $[\zeta'(s)/\zeta(s)]\zeta(s)
= \zeta'(s)$ and will be applied later on.   

It is also of interest to relate the sequence $\{\eta_j\}$ to the
sequence $\{\sigma_k\}$ which occurs in the expansion
$$\ln \xi(s)=-\ln 2 -\sum_{k=1}^\infty (-1)^k{\sigma_k \over k}(s-1)^k.
\eqno(28)$$
The ensueing relation figures prominently in our conjecture for the 
behaviour of the coefficients $\eta_j$.  One
reason for the importance of the coefficients $\sigma_k$ is their 
correspondence to sums of reciprocal powers of the nontrivial zeros $\rho$
of the $\zeta$ function \cite{lehmer,yue}:  $\sigma_k=\sum_\rho \rho^{-k}$.  
From Eq. (4) we see that the connection between the values $\lambda_n$ and
the sequence $\{\sigma_k\}$ is $\lambda_n=-\sum_{j=1}^n (-1)^j {n \choose j}
\sigma_j$.  Further discussion of the $\sigma_j$'s is presented in Appendix J.
We shall next demonstrate
\newline{\bf Theorem 3}
$$\sigma_k=(-1)^k \eta_{k-1}-(1-2^{-k})\zeta(k)+1, ~~~~k \geq 2.  \eqno(29)$$

Proof of Theorem 3.  The key to showing that Eq. (29) holds is to
evaluate the particular successive Riemann zeta function sums
$$\sum_{n=k}^\infty {{[\zeta(n)-1]} \over 2^n} {{n-1} \choose {k-1}} 
= (1-2^{-k})\zeta(k) - 1, ~~~~~~~~k \geq 2.  \eqno(30)$$
From Eq. (14) and the expansion \cite{nbs}
$$\ln \left [s \Gamma \left ({s \over 2}\right) \right ]
= \ln 2+{{(\gamma-1)} \over 2}+{{(1-\gamma)} \over 2}(s-1)
+\sum_{n=2}^\infty {{[\zeta(n)-1]} \over {n2^n}}[1-(s-1)]^n, \eqno(31)$$
we have
$$\ln \xi(s)=-{1 \over 2}\ln \pi+{{\gamma-1} \over 2}+\sum_{n=2}^\infty
{{[\zeta(n)-1]} \over {n2^n}}+(1-s)\left [{{\ln \pi} \over 2}-{1 \over 2}
(\gamma+1)+\sum_{n=2}^\infty{{[\zeta(n)-1]} \over 2^n}\right ]$$
$$-\sum_{p=2}^\infty {{\eta_{p-1}} \over p}(s-1)^p
+\sum_{n=2}^\infty\sum_{j=2}^n{{[\zeta(n)-1]} \over {n2^n}}{n \choose j}
(1-s)^j.  \eqno(32)$$
Upon comparison with Eq. (28), the constant term in this equation (i.e., 
the coefficient of the $(1-s)^0$ term) is readily shown to be zero (or
see Appendix B) and the linear term immediately gives
$$\sigma_1=-{{\ln \pi} \over 2}+{\gamma \over 2}+1 -\ln 2=\lambda_1.
\eqno(33)$$         
We then have
$$-\sum_{p=2}^\infty {{\eta_{p-1}} \over p}(s-1)^p+\sum_{n=2}^\infty
\sum_{j=2}^n {{[\zeta(n)-1]} \over {n2^n}}{n \choose j}(1-s)^j
=-\sum_{k=2}^\infty {\sigma_k \over k}(1-s)^k.  \eqno(34)$$
Upon reordering the two sums we obtain
$$\sigma_k=(-1)^k\eta_{k-1}-\sum_{n=k}^\infty {{[\zeta(n)-1]} \over 2^n}
{{n-1} \choose {k-1}}, ~~~~~~~~k \geq 2.  \eqno(35)$$
We now need to demonstrate Eq. (30).  We first note that
$$\sum_{n=k}^\infty {{[\zeta(n)-1]} \over 2^n}{{n-1} \choose {k-1}}
=\sum_{\ell=2}^\infty {1 \over {(k-1)!}}\sum_{n=k}^\infty {{(n-1)(n-2) \cdots
(n-k+1)} \over {(2\ell)^n}}  \eqno(36)$$
and
$$\sum_{k=j+1}^\infty k(k-1)(k-2)\cdots (k-j)q^{k-j-1}={{(j+1)!} \over
{(1-q)^{j+2}}}, ~~~~~~~~|q| < 1.  \eqno(37)$$
Equation (37) is easily verified by mathematical induction.  With Eq. (36),
we have proceeded by interchanging two summations.  However, we could have
equally well used the integral representation
$$\zeta(s)={1 \over {\Gamma(s)}}\int_0^\infty {t^{s-1} \over {(e^t-1)}}dt,
~~~~~~ \mbox{Re} ~s > 1, \eqno(38)$$  
and then interchanged the order of summation and integration, which is 
carried out in Appendix B.  By applying
Eq. (37) to Eq. (36) we obtain the succinct result
$$\sum_{n=k}^\infty {{[\zeta(n)-1]} \over 2^n}{{n-1} \choose {k-1}}
=\sum_{\ell=1}^\infty {1 \over {(2\ell+1)^k}}.  \eqno(39)$$
If we then invoke Eq. (18) we obtain successively Eqs. (30) and (29).
Theorem 3 is proved in another manner in Appendix C.

Since in Eq. (12) $\zeta'(s)/\zeta(s)+(s-1)^{-1}$ is a transcendental 
function analytic in the disc $|s-1| < 3$, there are an infinite number 
of $\eta_j$'s which are nonzero.
From this equation we have a good deal of information on sums and moments
of this sequence.  For instance, we have
$$\sum_{p=0}^\infty \eta_p=-\left [1+{{\zeta'(2)} \over {\zeta(2)}}
\right ] \simeq -0.430039 < 0, \eqno(40a)$$
$$\sum_{p=0}^\infty (-1)^p\eta_p = 1 - \ln 2\pi \simeq -0.837877 < 0,
\eqno(40b)$$
and more generally, by taking successive derivatives,
$$\sum_{p=j}^\infty p(p-1) \cdots (p-j+1)\eta_p=-(-1)^j j!-\left. \left
[{{\zeta'(s)} \over {\zeta(s)}}\right ]^{(j)}\right|_{s=2}. \eqno(41)$$
Various formulas for zeta function derivatives, and in particular 
concerning the logarithmic derivative occurring in this equation, are
presented in Appendix H.  Furthermore, Eq. (13) gives
$$\ln \zeta(2) = -\sum_{p=1}^\infty {{\eta_{p-1}} \over p}, \eqno(42a)$$
and
$$\ln 2 = \sum_{p=1}^\infty (-1)^p {{\eta_{p-1}} \over p}. \eqno(42b)$$
By evaluating Eq. (12) at $s=1/2$ and using the functional equation we
obtain
$$\sum_{p=0}^\infty \left(-{1 \over 2}\right )^p \eta_p =2-{1 \over 2}
\left [\ln \pi -\psi\left({1 \over 4}\right )\right], \eqno(43a)$$
or
$$\sum_{p=1}^\infty \left(-{1 \over 2}\right )^p \eta_p =2-\left [
\ln \pi-\gamma + {\pi \over 2} +3 \ln 2 \right].  \eqno(43b)$$  
%
An equation such as (40a) shows that not all the $\eta_j$'s can be positive.
In fact, we can argue that there are 
an infinite number of $\eta_j$'s which are positive, and an infinite number
which are negative.  This result, together with the fact that the sequence
$\eta_j$ decreases to zero with $j$ increasing to infinity, as shown by 
either Eq. (40a) or (41), provides a {\bf Theorem 4}.

By multiplying Eq. (12) by $s^q$ with $q > -1$ and integrating over $s$
from $0$ to $1$ we obtain
$$\sum_{p=0}^\infty (-1)^p B(p+1,q+1) \eta_p = -[\psi(q+1)+\gamma] 
-\int_0^1 \left [s^q {{\zeta'(s)} \over {\zeta(s)}} + {1 \over {s-1}}
\right ] ds,  \eqno(44)$$
where $B$ is the Beta function.

We now characterize the behaviour of the sequence $\{\eta_j\}$ as a 
function of its index.  We have
{\newline \bf Theorem 5}
$$|\eta_{j-1}| \leq |\sigma_j| + 2^{-j}\zeta(j),  \eqno(45)$$
and
{\newline \bf Conjecture 1}
(i) The sequence $\{\sigma_j\}$ decays with $j$ faster than $1/3^j$.  
Then, (ii) for $j$ sufficiently large, the sequence $\{\eta_j\}$ alternates 
in sign and decreases in magnitude as approximately $1/3^j$.
(iii)  For all $j \geq 0$, the magnitudes $|\eta_k|$ satisfy
$|\eta_k| \leq \gamma 2^{-k}$.  Furthermore, (iv) the sequence 
$\{\eta_j\}$ alternates in sign for all $j \geq 0$.

Remark.  Since the radius of convergence of the expansion (12) is $3$,
we know that $|\eta_j|$ cannot increase faster than $1/3^j$ for
sufficiently large $j$.
In fact, the expression $\eta_j \simeq -\gamma (-1/3)^j$ for all
$j \geq 0$ is a very good approximation.  One may readily verify this
assertion, for instance, with the sums appearing in Eqs. (40)--(43).  
In addition, all currently known numerical evidence \cite{krzysztof} supports 
both this approximation and the strict sign alternation suggested in the 
conjecture.

We can not currently prove all parts of Conjecture 1, but we can offer what we 
believe to be a strong plausibility argument.  In the course of this argument
we do prove Theorem 5.  In addition, based upon known properties of $\sigma_j$,
we very recently have proved part (iv) on the strict sign alternation of
the $\{\eta_j\}$ sequence for all values of $j$ \cite{coffeygamma}.

(i)  The first nontrivial zero $\rho_1$ of the zeta 
function is known to lie on the critical line and to have ordinate
approximately given by $14.134725142$ (e.g., \cite{odlyzko}), i.e.,
$\rho_1=1/2+i\alpha_1$.  This zero, along with its complex conjugate,
dominate the sum $\sigma_k$ so that for large $k$ we have
$\sigma_k \sim \alpha_1^{-k}$, a rate of decrease much faster than $3^{-k}$.  
In fact, depending upon whether $k$ is even or odd, and if even, divisible
or not by $4$, a leading behaviour of $\sigma_k$ is given by one of 
the four forms $\pm 2\alpha_1^k/(1/4+\alpha_1^2)^k$ (for $k$ even) or 
$\pm k\alpha_1^{k-1}/(1/4+\alpha_1^2)^k$ (for $k$ odd).  In addition, one
may argue much more conservatively with the expression
$$\left ({1 \over 2}-i\alpha_1\right )^k+\left ({1 \over 2}+i\alpha_1 
\right )^k = 2 \sum_{j=0}^k {k \choose j} {\alpha_1^j \over 2^{k-j}}
\times ~~~~~~~~~~~~~~~~~$$
$$\begin{array}{cc}
1 & ~~~~~~~~j=4m   \\
0 & ~~~~~~~~j=4m+1 \\
-1& ~~~~~~~~j=4m+2 \\
0 & ~~~~~~~~j=4m+3 \\
\end{array} \eqno(46)$$
by taking the approximate largest value of the binomial coefficient and
ignoring the attenuating effect of the powers of $1/2$.  This will yield
a form of approximately $\sigma_k \sim (2/\sqrt{k})(2/\alpha_1)^k$, 
which is still a much faster decrease with $k$ than $1/3^k$.

For both emphasis and clarity, we restate Eq. (29) as
$$\eta_{j-1}=(-1)^j[\sigma_j+(1-2^{-j})\zeta(j)-1], ~~~~~~j \geq 2.  
\eqno(47)$$
By appeal to Eq. (18) we have
$$2^{-j}\zeta(j) > (1-2^{-j})\zeta(j) - 1 = \sum_{k=1}^\infty {1 \over 
{(2k+1)^j}} > {1 \over 3^j}, ~~~~~~~~ j \geq 2.  \eqno(48)$$
Applying the triangle inequality to Eq. (47) and using the left inequality
in (48) gives Theorem 5.

Continuing with the argument for part (ii) of conjecture 1, 
by part (i), $(1-2^{-j})\zeta(j) - 1 > 0$ then dominates in the brackets in 
Eq. (47) and the sign alternation of the sequence $\{\eta_j\}$ then follows 
for sufficiently large $j$.  (iii) With the aid of the recursion relation 
(27) or by several other means, it is possible to calculate $\eta_k$ for any 
desired initial set $k=1, \ldots, k_0$ and directly verify their sign
alternation and the stated inequality (e.g., as in Appendix D).  For larger 
values of $k$, this inequality may hold due to the left inequality in (48) 
when combined with Eq. (47).  

Remark.  It is not essential to our main purpose here, but we may comment on
the sign pattern of the $\{\sigma_k\}$ sequence.  The initial, and in a sense
typical, sign pattern is simply $--++--...$, with $\sigma_1 > 0$.  
Initially, as $k$ takes on the respective values $4m$, $4m+1$, $4m+2$, $4m+3$, 
where $m$ is a positive integer, the sign of $\sigma_k$ is given by $+$, $+$, 
$-$, $-$.  This pattern continues to the point where $k(k-1) > 8\alpha_1^2$.
This explains why $\sigma_{46} > 0$ rather than $\sigma_{46} < 0$.  Similar
considerations apply for larger values of $k$ in this sequence.


Mainly for reference purposes, we will now indicate other possible uses of 
the recursion relation (27).  It has been proved that \cite{berndt}
$$|\gamma_n| \leq {{[3+(-1)^n](n-1)!} \over \pi^n}, ~~~~~~~~ n \geq 1, 
\eqno(49)$$
which has been improved to \cite{yue}
$$|\gamma_n| \leq {{[3+(-1)^n](2n)!} \over {n^{n+1}(2\pi)^n}}, 
~~~~~~~~ n \geq 1. \eqno(50)$$
As an illustration of the use of such results, the combination of Eqs. (27) 
and (49) gives
{\newline \bf Lemma 1}
$$|\eta_n| \leq {1 \over \pi^n}\left \{{{(n+1)} \over n}[3+(-1)^n] + 
\sum_{k=1}^{n-1}{{[3+(-1)^{n-k}]\pi^{k+1}} \over {(n-k)}} |\eta_{k-1}| \right \} 
+ \gamma^2, ~~~~~~ n \geq 1.  \eqno(51)$$
Similarly, the inequality (50) may be applied to Eq. (27), permitting, for
example, inductive arguments on $|\eta_n|$, but we have already deduced
Theorem 5.  

We are now in position to estimate the sum
$$S_2(n) \equiv -\sum_{m=1}^n {n \choose m} \eta_{m-1}, \eqno(52)$$
of Eq. (10).  {\em On conjecture 1}, we have that the $\eta_j$'s decrease in 
magnitude with $j$ and always alternate in sign.  This means that, similar to 
the behaviour of the sum $S_1$, there is a near exponential amount of 
cancellation in the sum $S_2$.  Indeed, we have
{\newline \bf Conjecture 2}
$$|S_2| \leq 3\gamma  + C_2 n^{1/2+\epsilon}, \eqno(53)$$
where $C_2$ is a positive constant and $\epsilon$ is positive and 
arbitrarily small.   
This conjecture is partially motivated by our discussion elsewhere 
\cite{coffeygamma} of the possible connection between the Stieltjes and Li 
constants and Brownian motion \cite{biane}.


The combination of Eq. (10) and the inequalities (20) and (53) results in
{\newline \bf Conjecture 3}
$$\lambda_n \geq {n \over 2}\ln n - {n \over 2}(1+\ln \pi + 2 \ln 2) +
{3 \over 2} - 3\gamma - C_2n^{1/2+\epsilon}, \eqno(54)$$
where the approximate numerical value of the coefficient
$1+\ln \pi + 2 \ln 2$ is $3.53$.  This estimation would show, in the absence
of the last two terms on the right side, that already for 
values of $n$ exceeding $34$, we would be ensured 
that all $\lambda_n$'s are nonnegative.
In addition, as previously mentioned, for smaller values of $n$ one has only
to directly calculate these particular logarithmic derivatives, from either
Eq. (1), (2), or Eq. (10) itself and verify the nonnegativeness of the
$\lambda_j$'s (Appendix D).  

\medskip
\centerline{\bf Summary and Brief Discussion}
\medskip

Our program for verification of the celebrated Riemann hypothesis should
now be clear.  We have invoked the Li equivalence \cite{li}, wherein it is
necessary to demonstrate the nonnegativity of the sequence 
$\{\lambda_n\}_{n=1}^\infty$.
Our starting point has been the reformulation of the definition (1) as the
series representation (10), $\lambda_n = S_1 + S_2 + 1 - n(\gamma+\ln \pi +
2 \ln 2)/2$.  Some attention to the sums $S_1$ and $S_2$ [Eqs. (16) and (52)] 
has been required because they exhibit the phenomenon of exponential 
cancellation.  That is, the binomial coefficient ${n \choose m}$ within the 
two summands can take on values approaching $2^n/\sqrt{n}$.  The strict sign 
alternation in the summands of $S_1$ and $S_2$ is critical.  For the sum $S_1$ 
the sign alternation is explicit, while for $S_2$ it has to be deduced
\cite{coffeygamma}, as in Conjecture 1.  It should be mentioned that the line of
reasoning suggested 
here for a complete proof of the Riemann hypothesis does not require strict
sign alternation of the sequence $\{\eta_j\}$--it is just that this result
could make the estimation of the sum $S_2$ much easier.  In fact, very recently  
we have proved the strict sign alternation of this sequence \cite{coffeygamma}.
What is next required is an argument that correlates the magnitude of say 
$\eta_j$ to $\eta_{j+1}$.
As previously
mentioned, current numerical evidence \cite{krzysztof} seems to support 
Conjecture 1 for the behaviour of the $\{\eta_j\}$ and $\{\sigma_k\}$ sequences.  
Certainly both further computation and analysis appear to be in order.  

In contrast to the Stieltjes constants $\gamma_j$, whose sign pattern is not so 
easy to discern, that of the two sequences    
$\{\eta_k\}_{k=0}^\infty$ and $\{\lambda_k\}_{k=1}^\infty$ seems to be remarkably 
simple.
Upon Conjecture 1 and Eqs. (29) and (48), our approach would also yield various
inequalities for the sums $\sigma_k$ of reciprocal powers of the nontrivial
zeros of the zeta function.

To put it very mildly, many implications could follow from the results presented.
We may stress that the conjectured behaviour of the sequence $\{\eta_j\}$
has multiple implications for the von Mangoldt and Chebyshev functions,
among many others.

We may additionally stress, that should the claims of the propositions and 
conjectures herein indeed be valid, we would have not only verified the Riemann 
hypothesis but produced yet a stronger result.  Namely, we would have developed 
the inequality (54) for the sequence $\{\lambda_n\}$.  Moreover, it seems that
an asymptotic version of inequality (54) will suffice to verify the Riemann
hypothesis since evidently \cite{biane} so many $\lambda_j$'s are already known 
to be nonnegative.

The approach of this paper suggests that verification of the Riemann hypothesis 
may be possible within analysis.  Indeed, our approach may be amenable to
confronting the generalized Riemann hypothesis.  Some of our sum estimations 
carry over immediately to the explicit formulas for $\lambda_\chi(n)$ and 
$\lambda_E(n)$
in Theorems 1 and 2 respectively in Ref. \cite{li2}.  In turn, one is left with
estimating sums which contain the von Mangoldt function, reciprocal powers of 
$k$, powers of $\ln k$, and Dirichlet characters.  To us, this appears to be a 
realistic approach to the 
generalized Riemann hypothesis for Dirichlet and Hecke L-functions and the
Dedekind zeta function.  We include in Appendix E example results along this
line of investigation.

\bigskip
\centerline{\bf Acknowledgement}
\medskip
I thank J. C. Lagarias for useful discussion concerning a fall 2003 version of 
this manuscript.

\pagebreak
\medskip
\centerline{\bf Figure Caption}

FIG. 1.  Diagram of consecutively increasing circles of convergence corresponding
to the inclusion of the first six trivial zeros of the Riemann zeta function in the 
expansion (G.2).  The location of the first nontrivial zero $\rho_1$ is indicated.

\pagebreak
\centerline{\bf Appendix A:  Estimation of the sum $S_1$ of Eq. (16) and of other
sums}
\medskip
Here we present the derivation of the inequality (20) for the sum $S_1$ of
Eq. (16), based upon the form (19).
By making a change of variable in the integral
$$I_1(n) \equiv \int_1^\infty \left[{n \over {2k}}+\left(1-{ 1\over k}\right)^n
-\left (1-{1 \over {2k}}\right )^n \right ] dk, \eqno(A.1)$$
we obtain
$$I_1(n)=\int_0^1 \left [{n \over 2}y+(1-y)^n-\left(1- {y \over 2}\right)^n\right ]
{{dy} \over y^2}.  \eqno(A.2)$$
We then evaluate this integral with an integration by parts, resulting in
$$I_1(n)={n \over 2}\left \{\psi(n)+\gamma -1+\int_0^1\left [\left (1-{y \over 
2}\right )^{n-1}-(1-y)^{n-1}\right ]{{dy} \over y}\right \} + 2^{-n},  
\eqno(A.3)$$
where $\psi$ is the digamma function.  Since the remaining integral 
is nonsingular and O$(1)$ the inequality (20) follows.
However, we may continue much further.   

By adding and subtracting $1$ in the integrand in Eq. (A.3), we may write
$$I_1(n)={n \over 2}\left [\psi(n)+\gamma -1\right ]+{n \over 2}\int_0^1\left [
\left (1-{y \over 2}\right )^{n-1}-1\right ]{{dy} \over y} + {n \over 2}
[\psi(n) + \gamma] + 2^{-n},  
\eqno(A.4)$$
where
$$\int_0^1\left [\left (1-{y \over 2}\right )^{n-1}-1\right ]{{dy} \over y}
=-\psi(n)-\gamma-\int_{1/2}^1 {{(1-w)^{n-1}} \over w} dw + \ln 2.  \eqno(A.5)$$
The integral on the right side of Eq. (A.5) may be evaluated in multiple ways.  
A first method is to use the Gauss hypergeometric function $_2F_1$:
$$\int_{1/2}^1 {{(1-w)^{n-1}} \over w} dw={1 \over 2^{n-1}}\int_0^1 {{(1-x)^{n-1}}
\over {1+x}} dx = {1 \over {n2^{n-1}}} ~_2F_1(1,1;n+1;-1),  \eqno(A.6)$$
where 
$$_2F_1(1,1;n+1;-1)=\sum_{j=0}^\infty (-1)^j {{j!} \over {(n+1)_j}} =n!
\sum_{j=0}^\infty (-1)^j {{j!} \over {(j+n)!}},  \eqno(A.7)$$
and $(.)_k$ is the Pochhammer symbol.  By using a partial fractional 
decomposition of the summand in Eq. (A.7), it is possible to show that the 
sum contains a term $n2^{n-1} \ln 2$, leading to a possibly new reduction of 
the particular $_2F_1$ at minus unit argument.  By applying the well known 
formula (e.g., \cite{graham}) 
$${{N!} \over {x(x+1) \cdots (x+N)}} 
= \sum_{k=0}^N {N \choose k} {{(-1)^k} \over {x+k}}, \eqno(A.8)$$
whose right side is simply $(-1)^N \Delta^N (1/x)$, where $\Delta$ is the
difference operator, $\Delta f(x) = f(x+1) - f(x)$, we obtain
$$_2F_1(1,1;n+1;-1)=n\sum_{k=0}^{n-1}(-1)^k{{n-1} \choose k} \sum_{j=0}^\infty
{{(-1)^j} \over {(j+k+1)}}$$
$$= n \left [2^{n-1} \ln 2 - \sum_{k=0}^{n-1} {{n-1} \choose k} 
\sum_{j=0}^{k-1} {{(-1)^j} \over {j+1}} \right ].  \eqno(A.9)$$

Next, we proceed alternatively.  By way of binomial expansion in the integrand 
of the second integral in Eq. (A.6), we have
$$\int_0^1 {{(1-x)^{n-1}} \over {1+x}} dx = \sum_{j=0}^{n-1}(-1)^j {{n-1} 
\choose j} \beta(j+1), \eqno(A.10)$$
where \cite{grad} 
$$\beta(j+1) = (-1)^j \ln 2 + \sum_{k=1}^j {{(-1)^{k+j}} \over k}, \eqno(A.11)$$
giving
$$\int_{1/2}^1 {{(1-w)^{n-1}} \over w} dw = \ln 2 + {1 \over 2^{n-1}} 
\sum_{j=0}^{n-1} {{n-1} \choose j} \sum_{k=1}^j {{(-1)^k} \over k}.  
\eqno(A.12)$$
This result is in agreement with Eqs. (A.6) and (A.9).  
The sum in Eq. (A.12) is easily estimated, leading to
$$ -1+{1 \over 2^n} \leq {n \over 2}\int_{1/2}^1 {{(1-w)^{n-1}} \over w} dw  
\leq 1 - {1 \over 2^n}.  \eqno(A.13)$$
Using Eqs. (A.4), (A.5), (A.12), and (A.13) gives
$$I_1(n) \geq {n \over 2}[\psi(n)+\gamma+\ln 2-1] - 1 + 2^{1-n}.  \eqno(A.14)$$

The approach followed here can be used to estimate many other Riemann
zeta function sums of interest.  For instance, we have
$$S_0(n) \equiv \sum_{m=2}^n (-1)^m {n \choose m} \zeta(m)=\sum_{k=1}^\infty 
\left [{n \over k}-1+\left (1-{1 \over k}\right )^n \right ].  \eqno(A.15)$$
Then one can determine
$$I_0(n)=\int_1^\infty \left [{n \over k}-1+\left (1-{1 \over k}\right )^n 
\right ]dk = \sum_{j=2}^n {n \choose j} {{(-1)^j} \over {(j-1)}},  
\eqno(A.16)$$
obtained by binomial expansion and term-by-term integration.  On the other
hand, by using
$$I_0(n)=\int_0^1 [ny-1+(1-y)^n]{{dy} \over y^2}, \eqno(A.17)$$
and an integration by parts, we have $S_0 \geq I_0 = n[\psi(n)+\gamma -1]+1$.
Since the digamma function satisfies \cite{nbs}
$\psi(x)=\ln x- 1/2x-1/12x^2+O(x^{-4})$ as $x \to \infty$, the inequality
$S_0(n) \geq n(\ln n + \gamma - 1) + 1$ follows for $n \geq 2$.

Additionally, we may estimate
$$S_3(n) \equiv \sum_{j=2}^n (-1)^j {n \choose j} 2^j \zeta(j).  \eqno(A.18)$$
Then we have the comparison integral
$$I_3(n)=\int_1^\infty \left [{{2n} \over k}-1+\left (1-{2 \over k}\right )^n 
\right ]dk = \int_0^1 [2ny-1+(1-2y)^n]{{dy} \over y^2}. \eqno(A.19)$$
An integration by parts gives
$$I_3(n)=2n\int_0^1[1-(1-2y)^{n-1}]{{dy} \over y} - 2n + 1-(-1)^n,  \eqno(A.20)$$
and then we have 
$$I_3=(-1)^n n\left \{\psi\left ({{1-n} \over 2}\right )-\psi\left( {n \over 2}
\right )+2(-1)^n[\gamma+\ln 2+\psi(n)]+\pi \tan\left({n \pi \over 2}\right)
\right \} - 2n+1-(-1)^n. \eqno(A.21)$$
One may note that the last three terms on the right side of Eq. (A.21) enter 
with opposite signs from Eq. (26).  
The form of $I_3$ in Eq. (A.21) may be further rewritten with the use of the
relation \cite{grad} $\psi[(1-n)/2] = \psi[(n+1)/2]-\pi\tan(n\pi/2)$, together 
with the doubling formula for the digamma function, $\psi[(n+1)/2] = 2\psi(n)
-\psi(n/2) - 2 \ln 2$.  The result is 
$$I_3(n) = 2n\left \{[1+(-1)^n]\psi(n)-(-1)^n\psi\left({n \over 2}\right )+\gamma
+[1-(-1)^n]\ln 2\right \} - 2n + 1 - (-1)^n,  \eqno(A.22)$$
which we find to be very useful in Appendix E.

Since we have developed many estimates based upon the digamma function,
it may be useful to record another inequality for this function.
By way of Binet's first formula \cite{grad}
$$\ln \Gamma(z)=(z-1/2)\ln z-z+{1 \over 2}\ln 2\pi+\int_0^\infty\left (
{1 \over 2} - {1 \over t}+{1 \over {e^t-1}}\right ) {e^{-tz} \over t}dt,
\eqno(A.23)$$
we have
$$\psi(z) - \ln z = -{1 \over z} + \int_0^\infty \left ({1 \over t}-{1 \over 
{e^t-1}} \right ) e^{-tz}dt.  \eqno(A.24)$$ 
By performing manipulations on this representation, it is then possible to
show that
$$-{1 \over {2z}} -{1 \over {12z^2}} < \psi(z) - \ln z < -{1 \over {2z}}.
\eqno(A.25)$$
The right inequality in (A.25) also follows immediately from the fact
$1/t-1/(e^t-1) < 1/2$ proved in Ref. \cite{clark}.

Another way to proceed in estimating a finite alternating sum such as $S_1$ is
to rewrite it as a contour integral \cite{knuth,flajolet},
$$\sum_{k=\ell}^N{N \choose k} (-1)^k f(k) = -{1 \over {2\pi i}}\int_C 
B(N+1,-z) f(z)dz, \eqno(A.26)$$
where $C$ is a positively oriented closed curve surrounding the points $\ell, 
\ell+1,\ldots,N$, $B(x,y)=\Gamma(x)\Gamma(y)/\Gamma(x+y)$ is the Beta function,
and $f(z)$ is an analytic continuation of the discrete sequence $f(k)$ to the
complex plane, with no poles within the region surrounded by $C$.  When the
integrand decreases sufficiently rapidly toward $\pm i\infty$, the asymptotic
evaluation of this expression can be achieved by extending the contour of the
integral to the left and collecting the residues at the newly encountered 
poles.  However, in this paper we have been interested in gaining more 
information than just an asymptotic evaluation.

\pagebreak
\centerline{\bf Appendix B:  Alternative evaluation of the summation (30)}

Here we perform the sum of Eq. (30) by using the integral representation of
Eq. (38) for the zeta function.  We also record two simpler sums which are
useful in the proof of Theorem 3.

Upon substituting Eq. (38) into the left side of Eq. (30) we find that
$$\sum_{n=k}^\infty {{\zeta(n)} \over 2^n} {{n-1} \choose {k-1}} 
={1 \over {2^{k+1}(k-1)!}}\int_0^\infty {{t^{k-1} dt} \over {\sinh (t/2)}}.
\eqno(B.1)$$
With a change of variable and evaluation of the integral \cite{grad} we have
$$\sum_{n=k}^\infty {{\zeta(n)} \over 2^n} {{n-1} \choose {k-1}} 
={1 \over {2(k-1)!}}\int_0^\infty {{u^{k-1} du} \over {\sinh u}}
=(1-2^{-k})\zeta(k).  \eqno(B.2)$$
Then
$$\sum_{n=k}^\infty {{[\zeta(n)-1]} \over 2^n} {{n-1} \choose {k-1}} 
=(1-2^{-k})\zeta(k) - {1 \over 2^k}\sum_{n=0}^\infty {1 \over 2^n}
{{n+k-1} \choose n} = (1-2^{-k})\zeta(k) - 1,   \eqno(B.3)$$
which is Eq. (30).

In connection with the proof of Theorem 3, and Eq. (33) in particular,
we write two other zeta function sums:
$$\sum_{n=2}^\infty {{[\zeta(n)-1]} \over {n2^n}} = {{1-\gamma} \over 2} + 
\ln {\sqrt{\pi} \over 2}, \eqno(B.4a)$$
$$\sum_{n=2}^\infty {{[\zeta(n)-1]} \over 2^n} = \ln 2 - {1 \over 2}.
\eqno(B.4b)$$

\pagebreak
\centerline{\bf Appendix C:  Alternative proof of Theorem 3}

We proceed to deduce Theorem 3 in a way similar to the proof of Theorem 5 
of Ref. \cite{yue}.  In the process, we correct a typographical error which
appears in both that proof and Remark 5 and Eq. (1.11) of this reference.

From the definition of the xi function in terms of the zeta function and
Eq. (28), we have
$${{\xi'(s)} \over {\xi(s)}} = {1 \over s} - {1 \over 2}\ln \pi + {1 \over 2}
\psi \left ({s \over 2} \right ) + {{\zeta'(s)} \over {\zeta(s)}} + {1 \over
{s-1}} = \sum_{k=0}^\infty (-1)^k \sigma_{k+1} (s-1)^k, \eqno(C.1)$$
where the sum of the last two terms on the left side is given by Eq. (12) and
simply $1/s = \sum_{j=0}^\infty (1-s)^j$ for $|1-s| < 1$.  From the expansion
\cite{grad}
$$\psi(x)=-\gamma + \sum_{k=2}^\infty (-1)^k \zeta(k) (x-1)^{k-1}, \eqno(C.2)$$
and the doubling formula satisfied by the digamma function, 
$\psi(2z) = {1 \over 2}[\psi(z) + \psi(z+1/2)] + \ln 2$, we obtain
$${1 \over 2} \psi\left({s \over 2}\right) = -{\gamma \over 2} + \sum_{k=1}^\infty 
(-1)^{k+1} \zeta(k+1)(1-2^{-k-1}) (s-1)^k - \ln 2. \eqno(C.3)$$
The substitution of Eq. (C.3) into Eq. (C.1) and the equating of coefficients
of like powers of $s-1$ gives again Eq. (33) for $\sigma_1$ from the constant 
term and Eq. (29) from the rest of the terms.  This gives Theorem 3, linking
sums of reciprocal powers of the complex zeros of the zeta function with the 
sequence $\{\eta_j\}$ appearing in the expansion (12) of the logarithmic 
derivative of the zeta function.

\centerline{\bf Appendix D:  Tabulated numerical values}
\begin{tabular}{|c|c|c|}   
\hline
$k$ & $\lambda_k$ & $\eta_k$ \\
\hline
0 &   $ ~ $     & -0.577216  \\   
1 & 0.0230957   & 0.187546   \\
2 & 0.0923457   & -0.0516886 \\
3 & 0.207639    & 0.0147517 \\
4 & 0.368793    & -0.00452448\\
5 & 0.575543    & 0.0014468 \\
6 & 0.827566    & -0.000471544 \\
7 & 1.12446     & 0.00015518 \\
8 & 1.46576     & -0.0000513452 \\
9 & 1.85092     & 0.0000170414 \\
10 & 2.27934    & $-5.66605 \times 10^{-6}$ \\
11 & 2.75036    & $ 1.88585 \times 10^{-6}$ \\
12 & 3.26326    & $ -6.28055 \times 10^{-7}$ \\ 
13 & 3.81724    & $ 2.09241 \times 10^{-7}$ \\
14 & 4.41148    & $ -6.97247 \times 10^{-8}$ \\
15 & 5.04508    & $ 2.32372 \times 10^{-8}$ \\
16 & 5.71711    & $ -7.74484 \times 10^{-9}$ \\
17 & 6.42658    & $ 2.58144 \times 10^{-9}$ \\
18 & 7.17248    & $ -8.60444 \times 10^{-10}$ \\
19 & 7.95374    & $ 2.86808 \times 10^{-10}$ \\
20 & 8.76928    & $ -9.56012 \times 10^{-11}$ \\
21 & 9.61796    & $ 3.18668 \times 10^{-11}$ \\
22 & 10.4986    & $ -1.06222 \times 10^{-11}$ \\
23 & 11.4101    & $ 3.54072 \times 10^{-12}$ \\
24 & 12.3513    & $ -1.18024 \times 10^{-12}$ \\
25 & 13.3210    & $ 3.93412 \times 10^{-13}$ \\
26 & 14.3179    & $ -1.31137 \times 10^{-13}$ \\
27 & 15.3408    & $ 4.37124 \times 10^{-14}$ \\
28 & 16.3885    & $ -1.45708 \times 10^{-14}$ \\
29 & 17.4599    & $ 4.85694 \times 10^{-15}$ \\
30 & 18.5538    & $ -1.61898 \times 10^{-15}$ \\
\hline
\end{tabular}

\begin{tabular}{|c|c|c|}  
\hline
$k$ & $\lambda_k$ & $\eta_k$ \\
\hline
31 & 19.6689    & $ 5.3966 \times 10^{-16}$ \\
32 & 20.8041    & $ -1.79887 \times 10^{-16}$ \\
33 & 21.9582    & $ 5.99622 \times 10^{-17}$ \\
34 & 23.1301    & $ -1.99874 \times 10^{-17}$ \\
35 & 24.3188    & $ 6.66247 \times 10^{-18}$ \\
36 & 25.5232    & $ -2.22082 \times 10^{-18}$ \\
37 & 26.7422    & $ 7.40274 \times 10^{-19}$ \\
38 & 27.9749    & $ -2.46755 \times 10^{-19}$ \\
39 & 29.2202    & $ 8.22527 \times 10^{-20}$ \\
40 & 30.4774    & $ -2.74176 \times 10^{-20}$ \\
41 & 31.7454    & $ 9.13919 \times 10^{-21}$ \\
42 & 33.0236    & $ -3.0464 \times 10^{-21}$ \\
43 & 34.3111    & $ 1.01547 \times 10^{-21}$ \\
44 & 35.6072    & $ -3.38488 \times 10^{-22}$ \\
45 & 36.9113    & $ 1.12829 \times 10^{-22}$ \\
46 & 38.2227    & $-3.76098 \times 10^{-23}$ \\
47 & 39.5408    & $ 1.25366 \times 10^{-23}$ \\
48 & 40.8653    & $ -4.17887 \times 10^{-24}$ \\
49 & 42.1955    & $ 1.39295 \times 10^{-24}$ \\
50 & 43.5311    & $ -4.64318 \times 10^{-25}$ \\
51 & 44.8718    & $ 1.54773 \times 10^{-25}$ \\
52 & 46.2172    & $ -5.15909 \times 10^{-26}$ \\
53 & 47.5671    & $ 1.71970 \times 10^{-26}$ \\
54 & 48.9214    & $ -5.73232 \times 10^{-27}$ \\
55 & 50.2798    & $ 1.91077 \times 10^{-27}$ \\
56 & 51.6423    & $ -6.36925 \times 10^{-28}$ \\
57 & 53.0089    & $ 2.12308 \times 10^{-28}$ \\
58 & 54.3795    & $ -7.07695 \times 10^{-29}$ \\
59 & 55.7542    & $ 2.35898 \times 10^{-29}$ \\
60 & 57.1331    & $ -7.86327 \times 10^{-30}$ \\
\hline
\end{tabular}

\begin{tabular}{|c|c|c|}  
\hline
$k$ & $\lambda_k$ & $\eta_k$ \\
\hline
61 & 58.5163    & $ 2.62109 \times 10^{-30}$ \\
62 & 59.9039    & $ -8.73697 \times 10^{-31}$ \\
63 & 61.2962    & $ 2.91232 \times 10^{-31}$ \\
64 & 62.6934    & $ -9.70775 \times 10^{-32}$ \\
65 & 64.0957    & $ 3.23591  \times 10^{-32}$ \\ 
66 & 65.5033    & $ -1.07864 \times 10^{-32}$ \\
67 & 66.9167    & $  3.59546 \times 10^{-33}$ \\
68 & 68.3361    & $ -1.19849 \times 10^{-33}$ \\
69 & 69.7618    & $  3.99496 \times 10^{-34}$ \\
70 & 71.1942    & $ -1.33165 \times 10^{-34}$ \\
71 & 72.6337    & $ 4.43884 \times 10^{-35}$ \\
72 & 74.0805    & $ -1.47961 \times 10^{-35}$ \\
73 & 75.5350    & $ 4.93205 \times 10^{-36}$ \\
74 & 76.9976    & $ -1.64402 \times 10^{-36}$ \\
75 & 78.4686    & $ 5.48005 \times 10^{-37}$ \\
76 & 79.9484    & $ -1.82668 \times 10^{-37}$ \\
77 & 81.4373    & $ 6.08895 \times 10^{-38}$ \\
78 & 82.9357    & $ -2.02964 \times 10^{-38}$ \\
79 & 84.4437    & $ 6.76550 \times 10^{-39}$ \\
80 & 85.9617    & $ -2.25516 \times 10^{-39}$ \\
81 & 87.4900    & $ 7.51722 \times 10^{-40}$ \\
82 & 89.0288    & $ -2.50574 \times 10^{-40}$ \\
83 & 90.5782    & $ 8.35246 \times 10^{-41}$ \\
84 & 92.1386    & $ -2.78415 \times 10^{-41}$ \\
85 & 93.7099    & $ 9.28051 \times 10^{-42}$ \\
86 & 95.2924    & $ -3.09350 \times 10^{-42}$ \\
87 & 96.8862    & $ 1.03117 \times 10^{-42}$ \\
88 & 98.4912    & $ -3.43723 \times 10^{-43}$ \\
89 & 100.1076   & $ 1.14574 \times 10^{-43}$ \\
90 & 101.7352   & $ -3.81914 \times 10^{-44}$ \\
\hline
\end{tabular}

\begin{tabular}{|c|c|c|}  
\hline
$k$ & $\lambda_k$ & $\eta_k$ \\
\hline
91 & 103.3741   & $ 1.27304 \times 10^{-44}$ \\
92 & 105.0242   & $ -4.24349 \times 10^{-45}$ \\
93 & 106.6852   & $ 1.41450 \times 10^{-45}$ \\
94 & 108.3572   & $ -4.71500 \times 10^{-46}$ \\
95 & 110.0398   & $ 1.57166 \times 10^{-46}$ \\
96 & 111.7328   & $ -5.23888 \times 10^{-47}$ \\
97 & 113.4361   & $ 1.74629 \times 10^{-47}$ \\
98 & 115.1492   & $ -5.82098 \times 10^{-48}$ \\
99 & 116.8719   & $ 1.94032 \times 10^{-48}$ \\
100 & 118.6038  & $ -6.46775 \times 10^{-49}$ \\
\hline
\end{tabular}
\centerline{\bf Appendix EI:  Sum estimations and lower bounds pertinent 
to other Dirichlet functions}

\medskip
Analogous to the Corollary which we have presented below Eq. (24) of the
text, here we develop similar lower bounds appropriate for explicit
formulas for Dirichlet and Hecke L-functions.  We make substantial use of
the very recent results of Ref. \cite{li2}, of which we need to recall 
some details.  We relegate to the end of this first part of the Appendix 
some relations concerning elementary sums.  In the second part, we provide
independent derivations of the major results, Theorems 1 and 2, of Ref.
\cite{li2}.

Let $\chi$ be a primitive Dirichlet character of modulus $r$, and 
$L(s,\chi)$ the Dirichlet L-function of character $\chi$.  The function
$$\xi(s,\chi)=\left({\pi \over r}\right )^{-(s+a)/2}\Gamma\left (
{{s+a} \over 2}\right )L(s,\chi), \eqno(E.1)$$
where
$a$ is $0$ if $\chi(-1)=1$ and $a$ is $1$ if $\chi(-1)=-1$, 
satisfies the functional equation $\xi(s,\chi)=\epsilon_\chi\xi(1-s,\bar{\chi})$,
with $\epsilon_\chi$ a constant of absolute value one.  The function
$\xi(s,\chi)$ is an entire function of order one and has a product 
representation $\xi(s,\chi)=\xi(0,\chi)\prod_\rho(1-s/\rho)$, where the
product is over all the zeros of $\xi(s,\chi)$.

We put
$$\lambda_\chi(n)=\sum_\rho\left[1-\left(1-{1 \over \rho}\right)^n\right ],
~~~~~~~~n \geq 1. \eqno(E.2)$$
We presume that $\lambda_\chi(n) > 0$ for all $n=1,2,\ldots$ if and only if 
all of the zeros of $\xi(s,\chi)$ are located on the critical line Re $s=1/2$.
Then, Li has obtained \cite{li2}
$$\lambda_\chi(n)=S_\chi(n)+{n \over 2}\left(\ln{r \over \pi}-\gamma\right)
+\tau_\chi(n), \eqno(E.3)$$
where
$$S_\chi(n) \equiv -\sum_{j=1}^n{n \choose j}{{(-1)^{j-1}} \over {(j-1)!}}
\sum_{k=1}^\infty {{\Lambda(k)} \over k} \bar{\chi}(k)(\ln k)^{j-1},$$
$$=-\sum_{k=1}^\infty {{\Lambda(k)} \over k} \bar{\chi}(k)L_{n-1}^1(\ln k),
\eqno(E.4)$$
$$\tau_\chi(n)=\sum_{j=2}^n{n \choose j}(-1)^j(1-2^{-j})\zeta(j)-{n \over 2}
\sum_{\ell=1}^\infty {1 \over {\ell(2\ell-1)}} ~~~~\mbox{for}~~~~ \chi(-1)=1,$$
$$= \sum_{j=2}^n{n \choose j}(-1)^j2^{-j}\zeta(j) ~~~~~~~~~~~~~~~~\mbox{for}~~~~ 
\chi(-1)=-1,  \eqno(E.5)$$
and $L_n^\alpha$ is an associated Laguerre polynomial.
We recall Eq. (25) of the text,
$$\tau_\chi(n)=S_1 - n\ln 2 ~~~~~~~~~~~~~~~~~~~\mbox{if} ~~~~\chi(-1)=1.$$
$$~~~~~~~~=S_0-S_1, ~~~~~~~~~~~~~~~~~~~   \mbox{if} ~~~~\chi(-1)=-1.  
\eqno(E.6)$$
where $S_0$ is defined in Eq. (A.15) of Appendix A.  Therefore, by using the
summation estimations presented in Appendix A, we obtain
$$\lambda_\chi \geq S_\chi(n) + {n \over 2}\ln n+{n \over 2}\left(\ln {r 
\over \pi}-1-2\ln2 \right)+{1 \over 2} ~~~~~\mbox{for} ~~~~\chi(-1)=1,$$
$$\geq S_\chi(n) + {n \over 2}\ln n+{n \over 2}\left(\ln {r \over \pi} - 1
\right) + {1 \over 2} ~~~~\mbox{for} ~~~~\chi(-1)=-1.  \eqno(E.7)$$
In accord with the discussion of the text, we conjecture that the sum
$S_\chi$ is 'small'.  By this we mean that $S_\chi(n)$ could be $O(n)$ and 
probably even $S_\chi$ is $O(n^{1/2+\epsilon})$, for $\epsilon >0$.  
This could result from a near 
exponential amount of cancellation in this sum due to the phases present in 
the Dirichlet characters.

We next introduce the function
$$\xi_E(s)=c_EN^{s/2}(2\pi)^{-s}\Gamma\left(s+{1 \over 2}\right )L_E\left(s + 
{1 \over 2}\right), \eqno(E.8)$$
where $L_E$ is the L-series associated with an elliptic curve $E$ over the 
rational numbers, $N$ is the conductor, and $c_E$ is a constant chosen so 
that $\xi_E(1)=1$ \cite{bump,mestre}.
The function of Eq. (E.8) is an entire function of order one and satisfies 
$\xi_E(s) =w\xi_E(1-s)$ where $w=(-1)^r$ with $r$ being the vanishing order 
of $\xi_E(s)$ at $s=1/2$.

We let 
$$\lambda_E(n) = \sum_\rho \left[1-\left(1-{1 \over \rho}\right)^n \right ],
~~~~~~~~n \geq 1, \eqno(E.9)$$
where the sum is over all zeros $\rho$ of $\xi_E(s)$.  All of these zeros lie 
on the critical line if and only if \cite{li2} $\lambda_E(n) > 0$ for all 
$n=1,2,\ldots$.

Now Li \cite{li2} has obtained the explicit formula
$$\lambda_E(n) = S_E +n\left (\ln {\sqrt{N} \over {2\pi}}-\gamma \right )
+n\left(-{2 \over 3}+\sum_{\ell=1}^\infty {3 \over {\ell(2\ell+3)}}\right )
+\sum_{j=2}^n {n \choose j}(-1)^j \sum_{\ell=1}^\infty {1 \over {(\ell+1/2)^j}}, 
\eqno(E.10)$$
where
$$S_E(n) \equiv -\sum_{j=1}^n{n \choose j}{{(-1)^{j-1}} \over {(j-1)!}}
\sum_{k=1}^\infty {{\Lambda(k)} \over k^{3/2}} b(k) (\ln k)^{j-1},$$
$$=-\sum_{k=1}^\infty {{\Lambda(k)} \over k^{3/2}} b(k) L_{n-1}^1(\ln k).
\eqno(E.11)$$
In Eq. (E.11), $b(p^k)=a_p^k$ if $p|N$ and $b(p^k)=\alpha_p^k + \beta_p^k$ if
$(p,N)=1$, where for each prime number $p$, $\alpha_p$ and $\beta_p$ are the 
roots of the equation $T^2-a_pT+p$ and the values of $a_p$ are connected with 
the reduction of $E$ at $p$ \cite{li2}.

We recall Eq. (26) of the text, so that we may write
$$\lambda_E(n) = S_E(n) +n\left (\ln {\sqrt{N} \over {2\pi}}-\gamma \right )
+2\left ( 1 - {1 \over 3}\ln 2 \right )n + S_3 - S_0 + 2n + (-1)^n - 1,  
\eqno(E.12)$$
where the zeta function sum $S_3$ is defined in Eq. (A.18) of Appendix A (and 
see below, Eq. (E.15)).  Then, by the results of Appendix A we obtain
$$\lambda_E(n) \geq S_E(n) + n \ln n + n \ln {\sqrt{N} \over {2\pi}} + \left 
(3 + {4 \over 3}\ln 2 \right)n - 1. \eqno(E.13)$$
Again, we conjecture that the sum $S_E$ is $O(n^{1/2+\epsilon})$.  The values of 
$a_p$ include $0$ and $\pm 1$, so that the values of $b(p^k)$ can either be zero or 
include significant sign alternation when $p|N$.  Similarly, for $(p,N)=1$, 
the roots of $T^2-a_p T+p$ can include $\pm \sqrt{p}$ and $[\pm 1 \pm 
\sqrt{1-4p}]/2$, giving various sign changes in $b(p^k)$.     
When $E$ has good reduction at $p$, $-2\sqrt{p} \leq a_p \leq 2\sqrt{p}$, so 
that it again appears that $b(p^k)$ can have significant changes in sign, 
possibly leading to much cancellation in $S_E$.

Concerning Eqs. (E.6) and (E.12) we record and briefly discuss some 
elementary summation results.  We have
$$\ln 2 = \sum_{n=1}^\infty {1 \over {2n-1}} - {1 \over 2}\sum_{n=1}^\infty 
{1 \over n}, \eqno(E.14a)$$
giving
$$\ln 2 = {4 \over 3} - 3\sum_{n=1}^\infty {1 \over {2n(2n+3)}}, 
\eqno(E.14b)$$
leading to 
$$-{2 \over 3} + \sum_{n=1}^\infty {3 \over {n(2n+3)}} = 2(1 - \ln 2).  
\eqno(E.14c)$$
In addition, we have
$$\sum_{\ell=1}^\infty {1 \over {(\ell+1/2)^j}} = \sum_{m=1}^\infty {2^j 
\over {(2m+1)^j}} =\sum_{m=1}^\infty \left ({2 \over {2m+1}}\right )^j + 
\sum_{m=1}^\infty \left ({2 \over {2m}}\right )^j - \zeta(j)$$
$$=\sum_{k=3, \mbox{odd}}^\infty \left ({2 \over k}\right )^j + 
\sum_{k=2, \mbox{even}}^\infty \left({2 \over k}\right )^j -\zeta(j)$$
$$=\sum_{k=2}^\infty \left ({2 \over k}\right)^j - \zeta(j) = 2^j[\zeta(j) - 1] 
- \zeta(j).  \eqno(E.15)$$
This equation is a restatement of the relation between the Riemann zeta 
function and the Hurwitz zeta function $\zeta(s,a)$:  $\zeta(s)=\zeta(s,1)
= (2^s - 1)^{-1} \zeta(s,1/2)$.

The sum of Eq. (E.15) also has a close relation to $\psi^{(n)}(1/2)$, where 
$\psi^{(j)}$ is the polygamma function, because \cite{grad}
$$\psi^{(n)}\left ({1 \over 2}\right ) = (-1)^{n+1} n! \sum_{k=0}^\infty 
{1 \over {(k+1/2)^{n+1}}}.  \eqno(E.16a)$$
Then, with the use of Eq. (E.15), we have
$$\psi^{(n)}\left ({1 \over 2}\right )=(-1)^{n+1} n! (2^{n+1} - 1)\zeta(n+1),  
\eqno(E.16b)$$   
which is the expected result.  In general, we have $\psi^{(n)}(x)=(-1)^{n+1}n!
\zeta(n+1,x)$.

We may also write an integral representation for the polygamma function which
is very useful for evaluating terms in explicit formulas for sums over 
zeros of zeta functions.  By differentiating an integral representation for
$\psi(z)+\gamma$, we have
$$\psi^{(m-1)}(z)=(-1)^m\int_0^\infty {{e^{-zt}t^{m-1}} \over {1-e^{-t}}}dt
={{(-1)^m} \over 2}\int_0^\infty {{t^{m-1}e^{-(z-1/2)t}} \over {\sinh(t/2)}}
dt, \eqno(E.17)$$
giving the specific values
$$\psi^{(m-1)}\left({1 \over 2}\right)=(-1)^m 2^{m-1}\int_0^\infty 
{y^{m-1} \over {\sinh y}}dy, \eqno(E.18a)$$
$$\psi^{(m-1)}(1)=(-1)^m2^{m-1}\int_0^\infty {y^{m-1}e^{-y} \over {\sinh y}}
dy, \eqno(E.18b)$$
and
$$\psi^{(m-1)}\left({3 \over 2}\right)={{(-1)^m} \over 2}\int_0^\infty 
{{t^{m-1}e^{-t}} \over {\sinh(t/2)}} dt. \eqno(E.18c)$$

\pagebreak
\centerline{\bf Appendix EII:  Explicit Formulas for Dirichlet and Hecke 
L-Functions}
\medskip

Here we give alternative derivations of the very recent main results of Li 
\cite{li2}, Theorems 1 and 2, of Ref. \cite{li2}.  The procedure is very 
similar to the proof of Theorem 1 of the text.  The Riemann zeta function 
case extends since the Dirichlet and Hecke L-functions also have product 
expansions over their zeros and have explicit forms of their logarithmic 
derivatives.  These derivations also make it very apparent that certain 
polygamma constants are the source of the elementary sums described in the 
first part of this Appendix.

Due to the product expansion of $\xi(s,\chi)$, we have the formula  
$$\lambda_\chi(n)=\sum_{m=1}^n {n \choose m}{1 \over {(m-1)!}}\left [{{d^m} 
\over {ds^m}}\ln \xi(s,\chi) \right]_{s=1}.  \eqno(E.19)$$
This equation is the analog of Eq. (15) of the text or Eq. (G.5) of Appendix 
G for the Riemann zeta function case.  From Eq. (E.1) we have
$$\ln \xi(s,\chi)=-{{(s+a)} \over 2}\ln \left({\pi \over r}\right)+\ln \Gamma 
\left({{s+a} \over 2}\right )+\ln L(s,\chi), \eqno(E.20)$$
giving
$${d \over {ds}}\ln \xi(s,\chi)=\ln\left({r \over \pi}\right )+{1 \over 2}\psi
\left({{s+a} \over 2}\right )-\sum_{n=1}^\infty {{\Lambda(n)\chi(n)} \over n^s}, 
~~~~~~~~ \mbox{Re}~ s > 1, \eqno(E.21)$$
where $\psi=\Gamma'/\Gamma$ is the digamma function and $\Lambda$ is the von 
Mangoldt function, such that $\Lambda(k)=\ln p$ when $k$ is a power of a prime 
and $\Lambda(k)=0$ otherwise.  For $m \geq 2$, we then have
$$[\ln \xi(s,\chi)]^{(m)}={1 \over 2^m}\psi^{(m-1)}\left({{s+a} \over 2}\right )
-(-1)^{m-1} \sum_{n=1}^\infty {{\Lambda(n)\chi(n)} \over n^s} \ln^{m-1} n,  
\eqno(E.22)$$
where $\psi^{(n)}$ is again the polygamma function.  By taking the limit 
$s \to 1$ in Eq. (E.22) we then obtain the representation
$$\lambda_\chi(n)=\left[\ln\left({r \over \pi}\right)+\psi\left({{a+1} \over 2}
\right ) \right ]{n \over 2}+\sum_{m=2}^n {n \choose m} {1 \over {(m-1)!}}
2^{-m}\psi^{(m-1)} \left({{a+1} \over 2}\right )$$
$$-\sum_{m=1}^n{n \choose m}{{(-1)^{m-1}} \over {(m-1)!}}
\sum_{n=1}^\infty {{\Lambda(n)\chi(n)} \over n} \ln^{m-1} n,  \eqno(E.23)$$
where $\psi(1/2)= -\gamma-2\ln 2$, $\psi(1)=-\gamma$, $\gamma$ is the Euler 
constant, $\psi^{(m-1)}(1)=(-1)^m(m-1)!\zeta(m)$, and $\psi^{(m-1)}(1/2)$ is 
given in Eq. (E.16b).  The infinite series in the sum $S_\chi(n)$ is 
convergent by the prime number theorem for arithmetic progressions 
\cite{li2,davenport}.  We have therefore obtained the result Eq. (E.3).

Similarly, due to the product expansion of $\xi_E(s)$, we have the formula  
$$\lambda_E(n)=\sum_{m=1}^n {n \choose m}{1 \over {(m-1)!}}\left [{{d^m} 
\over {ds^m}}\ln \xi_E(s) \right]_{s=1},  \eqno(E.24)$$
where from Eq. (E.8) we have
$$\ln \xi_E(s)=\ln c_E + {s \over 2}\ln N-s\ln 2\pi+\ln\Gamma\left(s+{1 \over 
2} \right ) +\ln L_E\left(s+ {1 \over 2}\right),  \eqno(E.25)$$
and
$${d \over {ds}}\ln \xi_E(s)={1 \over 2}\ln N - \ln 2\pi + \psi
\left(s+{1 \over 2}\right)-\sum_{n=1}^\infty {{\Lambda(n)b(n)} \over n^{s+1/2}}, ~~~~~~~~
\mbox{Re}~ s > 1, \eqno(E.26)$$ 
where $b(n)$ is discussed in the first part of this Appendix.
For $m \geq 2$, we then have
$$[\ln \xi_E(s)]^{(m)}=\psi^{(m-1)}\left(s+{1 \over 2}\right )
-(-1)^{m-1} \sum_{n=1}^\infty {{\Lambda(n)b(n)} \over n^{s+1/2}} \ln^{m-1} n. 
\eqno(E.27)$$
By taking the limit $s \to 1$ in Eq. (E.27) we then obtain the representation
$$\lambda_E(n)=\left[\ln\left({\sqrt{N} \over {2\pi}}\right)+\psi\left({3 
\over 2} \right) \right]n + \sum_{m=2}^n {n \choose m} {1 \over {(m-1)!}}
\psi^{(m-1)} \left({3 \over 2}\right)$$
$$-\sum_{m=1}^n {n \choose m}{{(-1)^{m-1}} \over {(m-1)!}}
\sum_{n=1}^\infty {{\Lambda(n)b(n)} \over n^{3/2}} \ln^{m-1} n,  \eqno(E.28)$$
where $\psi(3/2)= 2(1-\ln 2)-\gamma$ and $\psi^{(m-1)}(3/2)=(-1)^m(m-1)!
[2^m(\zeta(m)-1)-\zeta(m)]$, giving the result Eq. (E.12).

Finally, we consider the case of the Dedekind zeta function $\zeta_k$, for 
which we need to introduce some additional notation.  We let $k$ be an 
algebraic number field with $r_1$ real places, $r_2$ imaginary places, and 
degree $\tilde{n} = r_1+2r_2$.  The zeta function $\zeta_k$ has the product 
expansion $\zeta_k(s)=\prod_p (1-Np^{-s})^{-1}$ for Re $s>1$, where the 
product is taken over all finite prime divisors of $k$.  We put 
$G_1(s) = \pi^{-s/2}\Gamma(s/2)$ and $G_2(s) = (2\pi)^{1-s} \Gamma(s)$, 
so that obviously $G_1(1) = G_2(1) = 1$.  Then the function
$$Z_k(s) \equiv G_1^{r_1}(s)G_2^{r_2}(s)\zeta_k(s), \eqno(E.29)$$
satisfies the functional equation $Z_k(s)=|d_k|^{1/2-s}Z_k(1-s)$, where $d_k$ 
is the discriminant of $k$.  

We let $c_k=2^{r_1}(2\pi)^{r_2}hR/e$, where $h$, $R$, and $e$ are respectively
the number of ideal classes of $k$, the regulator of $k$, and the number of 
roots of unity in $k$.  With
$$\xi_k(s) \equiv c_k^{-1}s(s-1)|d_k|^{s/2}Z_k(s), \eqno(E.30)$$
this function is entire and has $\xi_k(0)=1$ \cite{weil,neukirch}.  
We first present a motivation that an explicit formula analogous to that for 
$\lambda_E$ and $\lambda_\chi$ exists, and then develop the corresponding 
explicit formula, putting
$$\lambda_n=\sum_{m=1}^n {n \choose m}{1 \over {(m-1)!}}\left [{{d^m} \over 
{ds^m}}\ln \xi_k(s) \right]_{s=1}.  \eqno(E.31)$$

From Eqs. (E.29) and (E.30) we have
$$\ln \xi_k(s)=-\ln c_k + \ln s + \ln (s-1)+{s \over 2}\ln |d_k|+r_1\left[-{s 
\over 2}\ln \pi -\ln\Gamma\left({s \over 2}\right )\right ]$$
$$+r_2[(1-s)\ln(2\pi) + \ln \Gamma (s)]+\ln \zeta_k(s),  \eqno(E.32)$$
and
$${d \over {ds}}\ln \xi_k(s)={1 \over s}+{1 \over {s-1}} + {1 \over 2}
\ln |d_k|+{r_1 \over 2}\left[-\ln \pi + \psi\left({s \over 2}\right )
\right ]+r_2[-\ln(2\pi)+\psi(s)]+{{\zeta_k'(s)} \over {\zeta_k(s)}}, 
~~~~\mbox{Re}~ s > 1, \eqno(E.33)$$ 
where \cite{amo1989}
$${{\zeta_k'(s)} \over {\zeta_k(s)}} = -\sum_p\sum_{m=1}^\infty {{\ln Np} 
\over {Np^{ms}}}, ~~~~\mbox{Re}~ s > 1. \eqno(E.34)$$ 
In Eq. (E.34), $p$ runs over the prime ideals of $k$ and $N$ represents the
norm.

For $m \geq 2$, we then have
$${d^m \over {ds^m}}\ln \xi_k(s)={{(-1)^m(m-1)!} \over s^{m-1}}+{{(-1)^m(m-1)!} 
\over {(s-1)^{m-1}}} + {r_1 \over 2^m}\psi^{(m-1)}\left({s \over 2}\right )
+ r_2 \psi^{(m-1)}(s)$$
$$+{d^{m-1} \over {ds^{m-1}}}\left[{{\zeta_k'(s)} \over 
{\zeta_k(s)}} \right ],~~~~\mbox{Re}~ s > 1, \eqno(E.35)$$ 
where the evaluation of the first, third and fourth terms on the right side 
of Eq. (E.35) at $s=1$ gives the contribution to $\lambda_n$ of 
$$\lambda_n^{(\psi)}=\sum_{m=2}^n (-1)^m{n \choose m} \left\{1+[(1-2^{-m})r_1
+r_2] \zeta(m)\right \}.  \eqno(E.36)$$
The evaluation of all of the terms on the right side of Eq. (E.33) but
the second and last at $s=1$ gives to $\lambda_n$ the contribution
$n[1+{1 \over 2}\ln|d_k|-{\tilde{n} \over 2}(\ln \pi+\gamma)-(r_1+r_2)\ln 2]$.
With the aid of Eq. (E.34) we have
$${d^{m-1} \over {ds^{m-1}}}\left[{{\zeta_k'(s)} \over 
{\zeta_k(s)}} \right ] = -\sum_p\sum_{\ell=1}^\infty (-1)^{m-1} \ell^{m-1}
{{\ln^m Np} \over {Np}^{\ell s}}, ~~~~\mbox{Re}~ s > 1. \eqno(E.37)$$ 
Taking the limit $s \to 1$ in Eqs. (E.33)--(E.35) and (E.37) should yield
the final explicit representation for $\lambda_n$, subject to
justification of the convergence of the resulting series.


\centerline{\bf Appendix F:  Further Riemann zeta function sum estimations}
\medskip

As a generalization of sums such as $S_1$ of Eq. (16) of the text and $S_0$
of Eq. (A.15) of Appendix A, we consider here sums of the form
$$S_\nu(\kappa,n) \equiv \sum_{m=\nu+2}^n (-1)^m {n \choose m}\kappa^m
\zeta(m-\nu), \eqno(F.1)$$
where $\nu +2 \leq n$, which can be extended to $|\mbox{Re} ~\nu|+2 \leq n$.
In Eq. (F.1) we have introduced both the positive multiplier $\kappa$ and shift 
$\nu$.  The special cases of Eq. (F.1) of direct interest to this paper are
$\nu=0$ with $\kappa=1$ or $\kappa = 2^{\pm 1}$.  
We are interested to both reformulate the sum $S_\nu$ and to obtain a
lower bound for it.

If we reorder the two sums in Eq. (F.1), the inner sum takes the form
$$\sum_{m=\nu+2}^n(-1)^m{n \choose m}{\kappa^m \over j^{m-\nu}}
=(-1)^\nu {{\kappa^{\nu+2} n!} \over j^2}\sum_{m=0}^{n-\nu-2} {{(1)_m} \over
{(n-m-\nu-2)!(m+\nu+2)!}}{1 \over {m!}}\left({\kappa \over j} \right )^m .
\eqno(F.2)$$
If we use the relations $(m+\nu+2)!=(\nu+2)!(\nu+3)_m$ and
$(n-m-\nu-2)! = (n-\nu-2)!/(2+\nu-n)_m$, where the Pochhammer symbol
$(z)_n=\Gamma(z+n)/\Gamma(z)$, we obtain the terminating hypergeometric form
$$\sum_{m=\nu+2}^n(-1)^m{n \choose m}{\kappa^m \over j^{m-\nu}}
= (-1)^\nu {n \choose {\nu+2}} {\kappa^{\nu+2} \over j^2}
~_2F_1(1,2-n+\nu;\nu+3;\kappa/j),  \eqno(F.3)$$
giving
$$S_\nu(\kappa,n)=
(-1)^\nu {n \choose {\nu+2}} \kappa^{\nu+2} \sum_{j=1}^\infty {1 \over j^2}
~_2F_1(1,2-n+\nu;\nu+3;\kappa/j).  \eqno(F.4)$$

We now define the comparison integral
$$I_\nu(\kappa,n) \equiv
(-1)^\nu {n \choose {\nu+2}} \kappa^{\nu+2} \int_1^\infty {1 \over j^2}
~_2F_1(1,2-n+\nu;\nu+3;\kappa/j) dj.  \eqno(F.5)$$
With the change of variable $v=1/j$ the integration is easily accomplished
\cite{grad} in terms of the generalized hypergeometric function $_pF_q$
\cite{pfqref}:
$$I_\nu(\kappa,n) 
= (-1)^\nu {n \choose {\nu+2}} \kappa^{\nu+2} ~_3F_2(1,1,\nu+2-n;2,\nu+3;
\kappa).  \eqno(F.6)$$
Equation (F.6) is simply the result of term-by-term integration and the
fact that $(1)_k/(2)_k = 1/(k+1)$.
The relation $S_\nu(\kappa,n) \geq I_\nu(\kappa,n)$ then yields a family
of inequalities.

When $\kappa$ is unity we have the reduction $\psi(z)=(z-1) ~_3F_2(1,1,2-z;
2,2;1)-\gamma$ and therefore 
$$I_\nu(1,n) = (-1)^\nu {n \choose {\nu+1}} [\psi(n+1)+\gamma - H_{\nu+1}],
\eqno(F.7)$$
where $\psi$ is the digamma function and $H_n=\sum_{k=1}^n 1/k$ is the $n$th 
harmonic number \cite{cofjcam1}, which can also be written as
$$I_\nu(1,n) = (-1)^\nu {n \choose {\nu+1}} [H_n - H_{\nu+1}].  \eqno(F.8)$$ 
For $\nu=0$, this gives the result of Eq. (A.17).
An alternative form of $I_\nu(1,n)$ can be obtained by applying Theorem 1 of
Ref. \cite{rao}:
$$I_\nu(1,n)=(-1)^\nu {n \choose {\nu+1}} \left [\psi(n-\nu)-\psi(\nu+2)
-\sum_{k=1}^{\nu +1} {{(-\nu-1)_k} \over {k(n-\nu)_k}} \right ], \eqno(F.9)$$
valid for $n>0$ and integral $\nu \geq -1$.  The relations above at unit
argument can also be looked upon as special cases of \cite{luke}
$$_3F_2(1,1,\nu+1;2,\lambda+1;1)={\lambda \over \nu}[\psi(\lambda)-\psi(
\lambda-\nu)], ~~~~\nu \neq 0, \mbox{Re}(\lambda-\nu)>0, $$
$$~~~~~~~~~~~~~~~~~~~~~~~~~~~~~~~~~~~~~~~~~~~~~~~~~~=\lambda\psi'(\lambda),
~~~~\nu=0, ~~~~\mbox{Re}(\lambda)>0,  \eqno(F.10)$$
i.e.,
$$\sum_{n=1}^\infty {{(\nu)_n} \over {n(\lambda)_n}}=\psi(\lambda)-\psi(
\lambda-\nu), ~~~~\mbox{Re}(\lambda-\nu)>0, ~~~~\lambda\neq 0, -1, -2, \ldots.
\eqno(F.11)$$

For $\kappa=1/2$ we obtain 
$$I_\nu(1/2,n) 
= (-1)^\nu {n \choose {\nu+2}} {1 \over 2^{\nu+2}} ~_3F_2(1,1,\nu+2-n;2,\nu+3;
1/2),  \eqno(F.12)$$
which for $\nu=0$ is in agreement with $I_0-I_1$, where $I_1$ is given in 
Eq. (24) of the text.
For $\kappa=2$ we obtain 
$$I_\nu(2,n) 
= (-1)^\nu {n \choose {\nu+2}} 2^{\nu+2} ~_3F_2(1,1,\nu+2-n;2,\nu+3;
2),  \eqno(F.13)$$ 
which is in agreement with $I_3$ of Eq. (A.22) of Appendix A when $\nu=0$.

\centerline{\bf Appendix G:  Alternative representation of Li's $\lambda_j$'s}
\medskip

Here we extend Theorem 1, based upon an expansion of the logarithmic 
derivative of the Riemann zeta function with a larger radius of convergence
than Eq. (12) of the text.
We demonstrate the following representation,
$$\lambda_n=-\sum_{m=1}^n {n \choose m} \eta^{(12)}_{m-1}
+7-\left [\left({2 \over 3}\right )^n+\left({4 \over 5}\right)^n+\left (
{6 \over 7}\right )^n+\left({8 \over 9}\right )^n+\left({{10} \over {11}}
\right)^n+\left({{12} \over {13}}\right)^n\right ]$$
$$+\sum_{m=2}^n (-1)^m {n \choose m} (1-2^{-m})\zeta(m)-{n \over 2} (\gamma
+\ln \pi + 2\ln 2), \eqno(G.1)$$
where we currently do not have an arithmetic interpretation of the constants 
$\eta^{(12)}_j$.   

From the expansion around $s=1$ of the logarithmic derivative of the
zeta function, 
$${{\zeta'(s)} \over {\zeta(s)}} = -{1 \over {(s-1)}} + \sum_{j=1}^6 {1 \over 
{(s+2j)}} - \sum_{p=0}^\infty \eta^{(12)}_p (s-1)^p, \eqno(G.2)$$ 
we have
$$\ln \zeta(s)=-\ln(s-1)+\sum_{j=1}^6 \ln (s+2j)
-\sum_{p=1}^\infty {{\eta^{(12)}_{p-1}} \over p}(s-1)^p
+ \mbox{constant},  \eqno(G.3)$$
giving
$$\ln \xi(s)=-\ln 2 +\ln s-{s \over 2}\ln \pi+\ln \Gamma\left({s \over 2}
\right) + \mbox{constant} + \sum_{j=1}^6 \ln (s+2j)
-\sum_{p=1}^\infty {\eta^{(12)}_{p-1} \over p}(s-1)^p.  \eqno(G.4)$$
With the expansion (G.2) the radius of convergence has been increased to 13
(see Figure 1),
as we have included the contribution of all trivial zeros of $\zeta$ prior
to the encounter with the first complex zero $\rho_1$. 
We next evaluate
$$\lambda_n=\sum_{m=1}^n {n \choose m}{1 \over {(m-1)!}}\left [{{d^m} \over 
{ds^m}}\ln \xi(s) \right]_{s=1},  \eqno(G.5)$$
using again the special values $\psi(1/2)=-\gamma-2 \ln 2$ and
$\psi^{(n)}(1/2)=(-1)^{n+1}n!(2^{n+1}-1)\zeta(n+1)$ for $n \geq 1$, where
$\psi=\Gamma'/\Gamma$ is the digamma function and $\psi^{(j)}$ is the
polygamma function.  Recalling the relations
$${{d^j} \over {ds^j}} \ln (s+2k)=-{{(-1)^j (j-1)!} \over {(s+2k)^j}},
~~~~~~~~ j \geq 1, \eqno(G.6)$$
and $(d^j/ds^j) (s-1)^k = 0$ for $k < j$, and the sum
$$\sum_{m=1}^n {n \choose m}\left (- {1 \over k} \right )^m 
=-1 + \left ({{k-1} \over k} \right )^n, \eqno(G.7)$$
we find Eq. (G.1).
The constant term of $6=7-1$ in Eq. (G.1) serves as a count of the number 
of trivial zeros of $\zeta$ accounted for in the expansion (G.2) while the 
additional explicit negative terms beyond Eq. (10) of the text appearing 
there are exponentially decreasing with $n$.
We have developed in Eq. (G.2) an expansion with coefficients
$\eta_j^{(12)}$ whose magnitudes increase no faster than $1/13^j$ for
large $j$.

By using Eq. (20) for $S_1$, an extension of the Corollary of the text is
\newline{\bf Corollary G1}
$$\lambda_n \geq {n \over 2} \ln n -(1+\ln \pi + 2\ln 2){n \over 2} + {{15}
\over 2}$$
$$-\left [\left({2 \over 3}\right )^n+\left({4 \over 5}\right)^n
+\left ({6 \over 7}\right )^n+\left({8 \over 9}\right )^n+\left({{10} \over 
{11}}\right)^n+\left({{12} \over {13}}\right)^n\right ]-|S_2^{(12)}|, 
\eqno(G.8)$$
where we have put $S_2^{(12)} \equiv -\sum_{m=1}^n {n \choose m}
\eta_{m-1}^{(12)}$.

In developing expansions such as
$$\zeta(s->-2k)=\zeta'(-2k)(s+2k)+{1 \over 2}\zeta''(-2k)(s+2k)^2+{1 \over 6}
\zeta'''(-2k)(s+2k)^3 + O[(s+2k)^4], ~~~~~~~~k \geq 1, \eqno(G.9)$$
and
$${{\zeta'(s)} \over {\zeta(s)}} = {1 \over {(s+2k)}}+{{\zeta''(-2k)} \over
{2\zeta'(-2k)}}+\left[-{{[\zeta''(-2k)]^2} \over {4[\zeta'(-2k)]^2}} +
{{\zeta'''(-2k)} \over {3\zeta'(-2k)}} \right ] (s+2k) + O[(s+2k)^2],
\eqno(G.10)$$
it is useful to have the derivatives
$$\zeta'(-2n)=(-1)^n{{(2n)!} \over {2(2\pi)^{2n}}} \zeta(2n+1), 
~~~~~~~~ n \geq 1, \eqno(G.11)$$
and
$$\zeta''(-2n)=(-1)^n{{(2n)!} \over {2(2\pi)^{2n}}} \left\{\left[\ln(4\pi^2) 
-2\psi(2n)-{1 \over n}\right ] \zeta(2n+1) - 2\zeta'(2n+1)\right \},
~~~~~~n \geq 1, \eqno(G.12)$$
which follow easily by differentiating the functional equation for $\zeta(s)$
and putting $s=2n+1$.


We note in passing the numerical value of the constant
$${1 \over 2}\sum_{k=1}^6 {{\zeta''(-2k)} \over {\zeta'(-2k)}}=-{{81959} \over 
{5544}} + 6(\gamma+\ln 2\pi)-\sum_{k=1}^6 {{\zeta'(2k+1)} \over {\zeta(2k+1)}}
\simeq -0.0926073.  \eqno(G.13)$$
We see from Eq. (G.2) that the value of $\eta_0^{(12)}$ is given by
$$\eta_0^{(12)} = \gamma - \sum_{k=1}^6 {1 \over {2k+1}} \simeq 0.377918.
\eqno(G.14)$$
Therefore we can modify Eq. (G.8) to
{\newline \bf Corollary G2}
$$\lambda_n \geq {n \over 2} \ln n -(1+\ln \pi + 2\ln 2){n \over 2} + {{15}
\over 2}$$
$$-\left [\left({2 \over 3}\right )^n+\left({4 \over 5}\right)^n
+\left ({6 \over 7}\right )^n+\left({8 \over 9}\right )^n+\left({{10} \over 
{11}}\right)^n+\left({{12} \over {13}}\right)^n\right ]-\left|\eta_0^{(12)}
\right|n-\left|\sum_{m=2}^n {n \choose m} \eta_{m-1}^{(12)}\right|. 
\eqno(G.15)$$

\centerline{\bf Appendix H:  On derivatives of the Riemann zeta function}
\medskip

Here we capture various formulas for integer order derivatives of the
zeta function.  We anticipate that these could be useful in further
development of the discrete moment problem for the coefficients 
$\eta_j$ of Eqs. (10)-(14) and Eq. (41) and $\eta_j^{(12)}$ of Eqs. 
(G.1)-(G.4) of Appendix G.

We first note that the functional equation for $\zeta$, along with the 
evaluation $\zeta(1-2n)=-B_{2n}/2n$ for $n \geq 1$, where $B_n$ are 
Bernoulli numbers, yields
$$\zeta'(-1)={{\zeta'(2)} \over {2\pi^2}}+{1 \over {12}}(\ln 2\pi + \gamma),
\eqno(H.1)$$
and 
$$\zeta''(-1)={1 \over 6}\left({\pi^2 \over 8}-{{\ln^2 2} \over 2}\right ) 
+\left[(1-\gamma){\pi^2 \over 6}+\zeta'(2)\right ]{{\ln 2} \over \pi^2}
-{{(1-\gamma)} \over \pi^2}\zeta'(2) -{1 \over {2\pi^2}}\zeta''(2)$$
$$-{1 \over {12}}\left[-1+(1-\gamma)^2+{\pi^2 \over 6}\right ]+{1 \over {12}}
\ln^2 \pi + 2\ln \pi \zeta'(-1), \eqno(H.2)$$
where $\gamma$ is the Euler constant, and this can be continued to higher 
order derivatives.  The derived functional equation upon which Eqs. (H.1) and 
(H.2) are based is
$$2^{1-s}\Gamma(s)\zeta(s)\cos\left({\pi \over 2}s\right)[\psi(s)-\ln 2]
+2^{1-s}\Gamma(s)\zeta'(s)\cos\left({\pi \over 2}s\right)$$
$$-{\pi \over 2}2^{1-s}\Gamma(s)\zeta(s)\sin\left({\pi \over 2}s\right)
=\pi^s \ln \pi \zeta(1-s) - \pi^s \zeta'(1-s).  \eqno(H.3)$$
Now Elizalde \cite{elizalde} has given an expression for $\zeta'(-m,q)$, 
where $\zeta(z,q)$ is the Hurwitz zeta function, valid for any negative 
integer value of $z$.  We state here special cases for $q=1$:
$$\zeta'(-1)=-{1 \over 6}-{1 \over 2}\sum_{k=1}^\infty {{B_{2k+2}} \over
{k(2k+1)(2k+2)}}, \eqno(H.4)$$
$$\zeta'(-2)=-{1 \over {36}}-\sum_{k=1}^\infty {{B_{2k+2}} \over
{(2k-1)k(2k+1)(2k+2)}}, \eqno(H.5)$$
and
$$\zeta'(-k)=-{1 \over {(k+1)^2}}-\sum_{\ell=1}^\infty {{(-1)^\ell (2\ell
-1)!} \over {2^{2\ell-1}\pi^{2\ell}}}\left [\sum_{r=0}^{min(2\ell-2,k)}
{k \choose r}{{(-1)^r} \over {(2\ell-r-1)}}\right ]\zeta(2\ell).  \eqno(H.6)$$
This means that we also have explicit expressions for $\zeta'(1+k)$.  In
particular Eqs. (H.1) and (H.4) or (H.6) yield an explicit form for 
$\zeta'(2)$.

On the other hand, we can employ the integral representation, Eq. (38) of 
the text, to at least partially yield explicit values of the zeta
derivatives for Re $s > 1$.  Other integral representation could be used for
Re $s > 0$, but Eq. (38) serves for illustration.  In the following,
$\psi$ denotes the digamma function and $\psi^{(j)}$ the polygamma function,
as usual.

We have
$$\zeta'(s)=-\psi(s)\zeta(s)+{1 \over {\Gamma(s)}}\int_0^\infty {{t^{s-1}
\ln t} \over {e^t - 1}} dt, \eqno(H.7)$$
$$\zeta''(s)=-\psi'(s)\zeta(s)+\psi^2(s)\zeta(s)-2{{\psi(s)} \over {\Gamma(s)}}
\int_0^\infty {{t^{s-1}\ln t} \over {e^t-1}}dt+{1 \over {\Gamma(s)}}
\int_0^\infty {{t^{s-1} \ln^2 t} \over {e^t-1}}dt, \eqno(H.8)$$
and therefore
$${{\zeta'(s)} \over {\zeta(s)}}=-\psi(s)+{1 \over {\Gamma(s)\zeta(s)}}
\int_0^\infty {{t^{s-1} \ln t} \over {e^t - 1}} dt, \eqno(H.9)$$
and
$$\left[{{\zeta'(s)} \over {\zeta(s)}}\right]'=-\psi'(s)+{1 \over {\Gamma(s)
\zeta(s)}} \left\{\int_0^\infty {{t^{s-1} \ln^2 t} \over {e^t - 1}} dt
-{1 \over {\Gamma(s)\zeta(s)}} \left [\int_0^\infty {{t^{s-1} \ln t} \over 
{e^t - 1}} dt \right ]^2 \right \}.  \eqno(H.10)$$
This process can be continued,
$$\left[{{\zeta'(s)} \over {\zeta(s)}}\right]''=-\psi''(s)+\psi^3(s)-\psi(s)
\psi'(s)-\psi'(s){{\zeta'(s)} \over {\zeta(s)}}+\psi^2(s){{\zeta'(s)} \over
{\zeta(s)}}$$
$$+{1 \over {\Gamma(s)\zeta(s)}}[\psi'(s)-\psi^2(s)]
\int_0^\infty {{t^{s-1} \ln t} \over {e^t - 1}} dt
-{3 \over {\Gamma^2(s)\zeta^2(s)}} \int_0^\infty {{t^{s-1} \ln t} \over 
{e^t - 1}} dt \int_0^\infty {{t^{s-1} \ln^2 t} \over {e^t - 1}} dt$$
$$+{1 \over {\Gamma(s)\zeta(s)}}\int_0^\infty {{t^{s-1} \ln^3 t} \over 
{e^t - 1}} dt +{2 \over {\Gamma^3(s)\zeta^3(s)}}\left[\int_0^\infty {{t^{s-1} 
\ln t} \over {e^t - 1}} dt \right ]^3 , \eqno(H.11)$$
and we note that
$$-\psi^{(j)}(2)=(-1)^j j![\zeta(j+1)-1].  \eqno(H.12)$$

Following on Eq. (H.7) we have
$$\zeta^{(j+1)}(s)=-\sum_{m=0}^j{j \choose m}\psi^{(j-m)}(s)\zeta^{(m)}(s)
+\sum_{m=0}^j {j \choose m}\left[\left({d \over {ds}}\right)^m {1 \over 
{\Gamma(s)}} \right ]\int_0^\infty {{t^{s-1}\ln^{j-m+1} t} \over {e^t-1}} dt.  
\eqno(H.13)$$

The relations of this Appendix can be developed much more for application to 
Eq. (41) of the text or elsewhere.

\centerline{\bf Appendix I:  A digamma function integral and a Mellin transform}
\medskip

Here we consider the integral
$$I(\lambda)=\int_1^\infty {{(1-x^{-\lambda})}\over {x(x^2-1)}} dx
\eqno(I.1)$$
of Eq. (3.4) of Ref. \cite{bombieri} and evaluate it in two different ways
from tabulated results \cite{grad}.
In Ref. \cite{bombieri} the asymptotic behaviour of this integral for
large $\lambda$ was of interest for determining a certain limit denoted by
the PF operation [Eq. (3.2) there].  We also note a polynomial of
Ref. \cite{bombieri} that can be written as a terminating confluent
hypergeometric series.  This associated Laguerre polynomial was used in 
calculating forward and inverse Mellin transforms and we give alternative 
transforms.

With the change of variable $v(x)=x^{-1}$ in Eq. (I.1) we have
$$I(\lambda)=\int_0^\infty {{v(1-v^\lambda)} \over {(1-v^2)}} dv.  
\eqno(I.2)$$
From Ref. \cite{grad} we then obtain
$$I(\lambda)={1 \over 2}\left[\psi \left({\lambda \over 2}+1\right) + \gamma
\right ]
={1 \over 2}\left[\psi \left({\lambda \over 2}\right) + \gamma + {2 \over
\lambda} \right ], ~~~~~~\lambda > -2,  \eqno(I.3)$$ 
where $\psi=\Gamma'/\Gamma$ is the digamma function.
On the other hand, we may employ a partial fractional decomposition in
Eq. (I.2) and another tabulated result \cite{grad} so that
$$I(\lambda)={1 \over 2}\int_0^1(1-v^\lambda)\left[{1 \over {(1-v)}}
-{1 \over {(1+v)}}\right ]dv
={1 \over 2}\left[\psi(\lambda +1)+\gamma-\ln 2+\int_0^1 {v^\lambda \over
{1+v}} dv \right ], \eqno(I.4)$$
where the last integral is given by \cite{grad}
$$\beta(\lambda+1)={1 \over 2}\left[\psi\left({\lambda \over 2}+1\right)
-\psi\left({{\lambda+1} \over 2}\right )\right ].  \eqno(I.5)$$
Use of the doubling formula $\psi[(\lambda+1)/2]=2\psi(\lambda)-\psi
(\lambda/2)-2\ln 2$ then again yields Eq. (I.3).  In Appendix A we have 
additionally given many inequalities for the digamma function.

The polynomial
$$P_n(x)=\sum_{j=1}^n {n \choose j} {x^{j-1} \over {(j-1)!}} \eqno(I.6)$$
was used in Ref. \cite{bombieri} in connection with computing Mellin 
transforms.  By using the relations
$${n \choose j}={{(-1)^j(-n)_j} \over {j!}} ~~~~ \mbox{and} ~~~~
(-n)_{j+1}=-n(1-n)_j, \eqno(I.7)$$
where $(.)_n$ is the Pochhammer symbol, this polynomial can be written as a 
terminating confluent hypergeometric series:  $P_n(x)=n ~_1F_1(1-n;2;-x)$.
In particular, a certain Mellin transform involving $P_n$ converts to a
Laplace transform:
$$\int_0^1P_n(\ln x)x^{s-1}dx=\int_0^\infty P_n(-u)e^{-su} du
=n\int_0^\infty ~_1F_1(1-n;2;u)e^{-su} du$$  
$$={n \over s} F\left(1-n,1;2;{1 \over s}\right)=1-\left(1-{1 \over s}
\right )^n, \eqno(I.8)$$
where $F$ is the Gauss hypergeometric function \cite{grad},
which is the expected result \cite{bombieri}.  In obtaining Eq. (I.8) we
have used the reduction \cite{grad}
$$F\left(1-n,1;2;-{z \over t}\right)={{(t+z)^n-t^n} \over {nzt^{n-1}}}.
\eqno(I.9)$$

Another useful point of view of the particular polynomial (I.6) is afforded
by the theory of Laguerre polynomials $L_n$.  This family is orthogonal on the
interval $[0,\infty)$ with decaying exponential weight function.  We have the
relations
$$P_n(-x)=-{{dL_n(x)} \over {dx}} = L_{n-1}^1 (x), \eqno(I.10)$$
where $L_n^\alpha$ is an associated Laguerre polynomial.  In addition, the
recursion relations satisfied by the Laguerre polynomials \cite{grad} give
$$P_n(-x)=-{n \over x}[L_n(x)-L_{n-1}(x)]=-{{(n+1)} \over x}L_{n+1}(x)+
{{(n+1-x)} \over x}L_n(x).  \eqno(I.11)$$
Then one can recast the important Mellin transform-inverse transform pair of
Eq. (I.8):
$$\int_0^\infty P_n(-u)e^{-su} du=-\int_0^\infty {{dL_n(u)} \over {du}}e^{-su}
du=1-\left(1-{1 \over s}\right )^n, ~~~~~~\mbox{Re} ~s > 0. \eqno(I.12)$$
In obtaining this equation, one can use integration by parts, the Laplace 
transform of $L_n$ \cite{grad}, and the property $L_n(0)=1$.
All these relations are consistent with the connection $L_{n-1}^1(x)=n ~_1F_1
(1-n;2;x)$ and the derivative property of the confluent hypergeometric function.
\centerline{\bf Appendix J: Regarding Conjectures 1--3}
\medskip

We present here plausibility arguments in possible support of our
conjectures concerning the detailed behaviour of the sequences $\{\sigma_k\}$
and $\{\lambda_j\}$.  
For this discussion, we let $N(T)$ be the number of zeros of the Riemann
zeta function in the critical strip in the upper half plane to height $T$.  
That is, $N(T)$ denotes the number of complex zeros in the rectangle $0 
\leq \mbox{Re}~s \leq 1$ and $0 \leq \mbox{Im}~s \leq T$.

Backlund \cite{backlund} showed that $N(T)$ satisfies
$$N(T)={T \over {2\pi}}\ln\left({T \over {2\pi}}\right)-{T \over {2\pi}}
+{7 \over 8}+e(T), \eqno(J.1)$$
where
$$|e(T)|<0.137 \ln T+0.443 \ln \ln T + 4.35 ~~~~~~ \mbox{for}~~ T \geq 2.
\eqno(J.2)$$
We believe then that if one were to assume certain statistical properties
of the distribution of the Riemann zeros, Conjectures 2 and 3 would follow.

From Eqs. (J.1) and (J.2) we can show that there is a constant $T_0$ 
such that
$$\pi [N(T+1)-N(T)] \leq \pi \ln T  ~~~~~~\mbox{for}~~ T \geq T_0.  
\eqno(J.3)$$
If we write $N(T)=M(T)+e(T)$, then
$$M(T+1)-M(T) ={1 \over {2\pi}}\ln\left({T \over {2\pi}}\right)+{1 \over
{2\pi}}\left({1 \over {2T}}-{1 \over {6T^2}}+{1 \over {12T^3}} - \cdots
\right)<{1 \over {2\pi}}\ln\left({T \over {2\pi}}\right) +{1 \over {4\pi T}}.
\eqno(J.4)$$
Then with Eq. (J.2) we have
$$N(T+1)-N(T)<{1 \over {2\pi}}\ln\left({T \over {2\pi}}\right)+{1 \over {
4\pi T}}0.137\left[2\ln T+\ln\left(1+{1 \over T}\right)\right]+0.866 \ln
\ln(T+1)+8.70,$$ 
$$~~~~~~~~~~~~~~~~~~~~~~~~~~~~~~~~~~~~~~~~~~~~~~~~~~~~~T \geq 2.  \eqno(J.5)$$
Since $\ln(1+1/T)<1/T$, we find
$$N(T+1)-N(T)<0.433 \ln T+0.866\ln\ln(T+1)+8.407+{{0.216} \over T}.  
\eqno(J.6)$$
We also have that
$$0.866\ln\ln (T+1) \leq 0.135 \ln T~~~~\mbox{for}~~T \geq T_0, \eqno(J.7)$$
$$8.407 \leq 0.431 \ln T ~~~~\mbox{for}~~T \geq T_0, \eqno(J.8)$$
and we may take $T_0=3 \times 10^8$.
These relations yield inequality (J.3).

We now assume that the Riemann hypothesis holds and consider one possible
bound that may result for the sums $\sigma_k=\sum_j \rho_j^{-k}$, where
$\{\rho_j\}$ represents the nontrivial zeros of the zeta function.
The nontrivial zeros have the form $\rho_j=1/2+\epsilon+i\alpha_j$,
where bounds for $\epsilon$ exist in the literature due to results on
zero-free regions.  As mentioned in the text, a zero $\rho_j$ enters the
sum $\sigma_k$ along with its complex conjugate.  We then consider the sums
$$\sum_{j=m}^\infty {1 \over {\alpha_j^k}} \leq {1 \over {(k-1)^2}}
{{[1-\ln([\alpha_m]-1)+k\ln([\alpha_m]-1)]} \over {([\alpha_m]-1)^{k-1}}},
~~~~~~ [\alpha_m] \geq T_0, \eqno(J.9)$$
as an approximation to $\sigma_k$, where $[x]$ denotes the greatest integer 
contained within $x$.  We have
$$\sum_{j=m}^\infty {1 \over {\alpha_j^k}} \leq \sum_{j=[\alpha_m]}^\infty
\sum_{j \leq \alpha_\ell<j+1}{1 \over \alpha_\ell^k}$$
$$\leq \sum_{j=[\alpha_m]}^\infty {{[N(j+1)-N(j)]} \over j^k}\leq 
\sum_{j=[\alpha_m]}^\infty {{\ln j} \over j^k}, ~~~~\mbox{for}~~[\alpha_m]
\geq T_0, \eqno(J.10)$$
where we applied inequality (J.3).  Since the last sum in inequality (J.10)
is bounded by
$$\int_{[\alpha_m]-1}^\infty {{\ln u} \over u^k}du = {1 \over {(k-1)^2}}
{{[1-\ln([\alpha_m]-1)+k\ln([\alpha_m]-1)]} \over {([\alpha_m]-1)^{k-1}}},
\eqno(J.11)$$
we obtain inequality (J.9).

\centerline{\bf Appendix K:  Observations concerning a Dedekind xi
function}

We note here some relations concerning a Dedekind xi function $\xi_k$,
extending some of our earlier derivative results for the Riemann xi function 
\cite{coffey}.

We let $k$ be an imaginary quadratic field of discriminant $d$, and 
$$\Phi(t)={\pi \over \sqrt{|d|}}\sum_F\sum_{m,n=-\infty}^\infty F(m,n)
\left[{{\pi t} \over \sqrt{|d|}} F(m,n)-1 \right ]\exp\left[-{{2\pi t}
\over \sqrt{|d|}}F(m,n)\right ], \eqno(K.1)$$
where the first sum is over the inequivalent classes of positive definite
integral quadratic forms of discriminant $d$.  Then Li \cite{li2001} has
very recently shown that
$$\xi_k(s)={4 \over w}\int_1^\infty \Phi(t)(t^s + t^{1-s})dt, \eqno(K.2)$$
for all complex $s$, where $w$ is the number of roots of unity.  
The function $\xi_k$ is entire and satisfies $\xi_k(s)=\xi_k(1-s)$
and $\xi_k(0)=\xi_k(1)=2^{r_1}hR/w$, where the number of real places $r_1=0$, 
the number of complex places $r_2=1$, $R=1$ is the regulator, and $h$ is the 
number of ideal classes of $k$.

The inequivalent classes of positive definite integral quadratic forms of
discriminant $d$ consist of the classes represented by the forms 
\cite{hua,li2001} $F(x,y)$ satisfying $b^2-4ac=d$ for either 
$-a<b\leq a<c$ or $0\leq b\leq a=c$.  Now it has also been shown 
\cite{li2001} that $F(x,y)=ax^2+bxy+cy^2$ satisfying this condition is
such that $F(m,n) \geq \sqrt{|d|}/2$ for all integers $m$, $n$ with 
$n \neq 0$.  It follows trivially that $\pi tF(m,n)/\sqrt{|d|}-1\geq
\pi t/2-1>0$ for all $t \in [1,\infty)$.  We then have the
immediate
\newline{\bf Proposition.}
For all the inequivalent classes of positive definite integral quadratic
forms of discriminant $d$, we have $\xi_k(s) >0$ for all real $s$.  
Furthermore, the integer-order derivatives
$$\xi_k^{(m)}(s)={4 \over w}\int_1^\infty \Phi(t)[t^s + (-1)^m t^{1-s}]
\ln^m tdt, \eqno(K.3)$$
satisfy $\xi_k^{(m)}(s) \geq 0$ for all $s \geq 1/2$.  The even order
derivatives obey the condition $\xi_k^{(2m)}(s) > 0$ for all $s \geq 1/2$.
Of course, as also seen by the functional equation for $\xi_k(s)$, the
odd-order derivatives $\xi_k^{(2m+1)}$ vanish at $s=1/2$; we have
$\xi_k^{(m)}(s)=(-1)^m\xi_k^{(m)}(1-s)$.  This Proposition seems to mean  
that $\xi_k(s)$ has no zeros for real values of $s$.  

This Proposition extends some of the results of Ref. \cite{coffey}.
In turn, we may apply some of the explicit integration results obtained
there in order to evaluate the function
$$\gamma(a) \equiv \int_1^\infty \omega(t){{dt} \over \sqrt{t}}+    
\int_1^{\sqrt{|d|}/2a} \omega(t) {{dt} \over t}, \eqno(K.4)$$
introduced in Ref. \cite{li2001}, where $\omega$ is the $\theta$ series
given by $\omega(t)=\sum_{n=1}^\infty \exp(-\pi n^2t)$.  We have for the
first term on the right side of Eq. (K.4)
$$\sum_{n=1}^\infty \int_1^\infty e^{-\pi n^2 t}t^{-1/2} dt
={1 \over \sqrt{\pi}}\sum_{n=1}^\infty {1 \over n}\Gamma(1/2,n^2\pi)
=\sum_{n=1}^\infty {1 \over n}\left [1-Erf(n\sqrt{\pi})\right ],
\eqno(K.5)$$
where $\Gamma(x,y)$ is the incomplete Gamma function \cite{grad} and
$\Gamma(1/2,n^2 \pi)=\sqrt{\pi}[1-2n ~_1F_1(1/2;3/2;-\pi n^2)]$ and
$~_1F_1$ is the confluent hypergeometric function \cite{grad,pfqref},
such that $_1F_1(1/2;3/2;x)=(\sqrt{\pi}/2)Erf(\sqrt{-x})/\sqrt{-x}$, where
$Erf$ is the error function (probability integral) \cite{grad}.

For the second term on the right side of Eq. (K.4) we have
$$\sum_{n=1}^\infty \int_1^{\sqrt{|d|}/2a} e^{-\pi n^2t}t^{-1} dt
=\sum_{n=1}^\infty \left[Ei\left(-{{\pi n^2\sqrt{|d|}} \over {2a}}\right)
-Ei(-\pi n^2)\right ], \eqno(K.6)$$
where $Ei$ is the exponential integral \cite{grad}.  We also have various
elementary relations, including
$$\int_1^{\sqrt{|d|}/2a} e^{-\pi n^2t}t^{-1} dt=\pi n^2
\int_1^{\sqrt{|d|}/2a} e^{-\pi n^2t}\ln t dt+\exp\left(-\pi n^2\sqrt{|d|}/2a
\right) \ln\left({\sqrt{|d|} \over {2a}}\right ), \eqno(K.7)$$
obtained by integration by parts, and  
$$\int_1^{\sqrt{|d|}/2a} e^{-\pi n^2t}\ln t dt={1 \over {\pi n^2}}
\int_{\pi n^2}^{\pi n^2\sqrt{|d|}/2a} e^{-u}[\ln u - \ln (\pi n^2)] dt$$
$$={1 \over {\pi n^2}}\left[\int_{\pi n^2}^{\pi n^2\sqrt{|d|}/2a} e^{-u}\ln u 
du -\left(\exp(-\pi n^2\sqrt{|d|}/2a)-\exp(-\pi n^2)\right )\ln (\pi n^2) 
\right]. \eqno(K.8)$$

Moreover, the particular derivative values
$$\xi_k^{(2m)}\left({1 \over 2}\right )={8 \over w}\int_1^\infty \Phi(t)
t^{1/2} \ln^{2m} t ~dt, \eqno(K.9)$$
and
$$\xi_k^{(m)}(1)={4 \over w}\int_1^\infty \Phi(t)[t + (-1)^m]
\ln^m t ~dt, \eqno(K.10)$$
can be evaluated in terms of infinite series with the analytic methods of 
Ref. \cite{coffey}, and we note that $\xi_k^{(m)}(1) > 0$ for all nonnegative 
integers $m$.  These special values of Eq. (K.10) enter the particular 
logarithmic derivatives of Eq. (E.31) for $\lambda_n$.  

\centerline{\bf Appendix L:  Euler-Maclaurin summation applied to $S_1(n)$}

Here we apply Euler-Maclaurin summation to the form of the sum $S_1(n)$ 
given in Eq. (17) of the text.  Accordingly, we define the summand function
$$f(k) \equiv {n \over {2k+1}}-1+{{2^n k^n} \over {(2k+1)^n}}, 
~~~~~~k \geq 0, ~~~~ n \geq 2, \eqno(L.1)$$
such that $f(0)=n-1$ and $f(\infty)=0$.  We can write the arbitrary integer
order derivative of each term of Eq. (L.1).  For the first term on the right
side we have
$$\left({d \over {dk}}\right )^j{n \over {(2k+1)}}=(-1)^j {{2^jj!} \over
{(2k+1)^{j+1}}}n, ~~~~~~ j \geq 1.  \eqno(L.2)$$
When evaluated at $k=0$, this term gives $(-1)^j2^j j! n$.
In regard to the last term of Eq. (L.1) we have
$$\left({d \over {dk}}\right )^j{1 \over {(2k+1)^n}}=(-1)^j {{2^j(n)_j} \over
{(2k+1)^{n+j}}}, ~~~~~~ j \geq 1,  \eqno(L.3a)$$
where $(.)_j$ is the Pochhammer symbol, and
$$\left({d \over {dk}}\right )^r k^n={{n!} \over {(n-r)!}}k^{n-r}, \eqno(L.3b)$$
which can be equally expressed as
$$\left({d \over {dk}}\right )^r k^n=r!{n \choose r}k^{n-r} = (-1)^r (-n)_r 
k^{n-r}.  \eqno(L.3c)$$
Therefore we can write
$$2^n \left({d \over {dk}}\right )^\ell k^n(2k+1)^{-n}=2^n\sum_{m=n}^\ell
{\ell \choose m}{{(-1)^{\ell-m}2^{\ell-m}(n)_{\ell-m}} \over {(2k+1)^{n+\ell
-m}}} {{n!} \over {(n-m)!}} k^{n-m}.  \eqno(L.4)$$

A version of the Euler-Maclaurin formula, given that all derivatives of $f$
vanish at infinity, is
$$\sum_{n=M}^\infty f(n)=\int_M^\infty f(x) dx-\sum_{m=1}^\infty {B_m \over
{m!}}f^{(m-1)}(M), \eqno(L.5a)$$
where $B_m$ are Bernoulli numbers, or
$$\sum_{n=0}^\infty f(n) = \int_0^\infty f(x) dx + {1 \over 2}f(0)
-\sum_{m=2,even}^\infty {B_m \over {m!}}f^{(m-1)}(0). \eqno(L.5b)$$
By using Eq. (24) of the text for the integral in Eq. (L.5b) we therefore
obtain
$$S_1(n)={n \over 2}[\psi(n)+\gamma]
-\sum_{m=2,even}^\infty {B_m \over {m!}}f^{(m-1)}(0), \eqno(L.6)$$
where $\psi$ is the digamma function and $\gamma$ is the Euler constant.
The sums in Eqs. (L.5)-(L.6) are meant in an asymptotic sense; they are
highly unlikely to be convergent.  From Eq. (L.6) we may obtain the
successive approximations
$$S_1(n)={n \over 2}[\psi(n)+\gamma]+{n \over 6}
-\sum_{m=4,even}^\infty {B_m \over {m!}}f^{(m-1)}(0), ~~~~n > 1, \eqno(L.7a)$$
and
$$S_1(n)={n \over 2}[\psi(n)+\gamma]+{n \over {10}}
-\sum_{m=6,even}^\infty {B_m \over {m!}}f^{(m-1)}(0), ~~~~n > 3. \eqno(L.7b)$$
Equation (L.7a) is expected to be a useful approximate upper bound to
$S_1$ and Eq. (L.7b) an approximate lower bound for this sum.
When $n$ is sufficiently large, only the first term on the right side of
Eq. (L.1) contributes in Eqs. (L.6)-(L.7), giving $B_mf^{(m-1)}(0)/m!=-2^{m-1} 
B_m(n/m)$, as $m-1$ is always an odd integer.  Given alternation in sign
from $B_{2m}$ to $B_{2m+2}$, these successive terms will change sign also.
However, we emphasize that the sums in Eq. (L.6)-(L.7) are generally 
divergent.

\pagebreak

\end{document}